\documentclass[12pt]{iopart}

\usepackage{graphicx}
\usepackage{pgfplots}
\usepackage{caption}
\usepackage{subcaption}
\usepackage{layouts}
\usepackage{subcaption}
\usepackage{setstack}
\usepackage{amssymb}
\usepackage{appendix}
\usepackage{braket}
\usepackage{hyperref}
\usepackage{epstopdf, epsfig}
\usepackage{algorithm}
\usepackage{algpseudocode}
\usepackage{ifthen}
\usepackage{iopams}
\usepackage{siunitx}
\usepackage{tikz}
\usetikzlibrary{quantikz2}
\usetikzlibrary{fit}
\usetikzlibrary{shapes,arrows}
\usetikzlibrary{shapes.geometric, arrows}

\newcommand{\step}[1]{\item \textbf{#1}}

\begin{document}

\title[State Preparation for bell-shaped probability distribution]{Quantum state preparation for bell-shaped probability distributions using deconvolution methods}

\author{Kiratholly Nandakumar Madhav Sharma$^{1,2}$, Camille de Valk$^2$, Ankur Raina$^3$, Julian van Velzen$^2$}

\address{$^1$Department of Physics, Indian Institute of Science Education and Research Bhopal}
\address{$^2$Capgemini Quantum Lab}
\address{$^3$Department of EECS, Indian Institute of Science Education and Research Bhopal}
\ead{ankur@iiserb.ac.in}
\vspace{10pt}
\begin{indented}
\item[]May 2024
\end{indented}

\begin{abstract}
Quantum systems are a natural choice for generating probability distributions due to the phenomena of quantum measurements. 
The data that we observe in nature from various physical phenomena can be modelled using quantum circuits.
To load this data, which is mostly in the form of a probability distribution, we present a hybrid classical-quantum approach.  
The classical pre-processing step is based on the concept of deconvolution of discrete signals. 
We use the Jensen-Shannon distance as the cost function to quantify the closeness of the outcome from the classical step and the target distribution. 
The chosen cost function is symmetric and allows us to perform the deconvolution step using any appropriate optimization algorithm. 
The output from the deconvolution step is used to construct the quantum circuit required to load the given probability distribution, leading to an overall reduction in circuit depth.
The deconvolution step splits a bell-shaped probability mass function into smaller probability mass functions, and this paves the way for parallel data processing in quantum hardware, which consists of a quantum adder circuit as the penultimate step before measurement. 
We tested the algorithm on IBM Quantum simulators and on the IBMQ Kolkata quantum computer, having a 27-qubit quantum processor.
We validated the hybrid Classical-Quantum algorithm by loading two different distributions of bell shape. 
Specifically, we loaded 7 and 15-element PMF for (i) Standard Normal distribution and (ii) Laplace distribution. 
\end{abstract}

%
%
%
%
%

\section{Introduction}
Traditional methods to model various real-world
phenomena use random processes or random variables whose distribution is to be learnt.
In classical computing machines or systems, the process of learning the distribution of a random source is termed stochastic modelling.
Or, the problem can be alternatively seen as generating a given probability distribution, namely the target distribution. 
In such cases, it becomes essential to learn the parameters of that distribution to generate samples from it easily. 
Seen from a quantum computing lens, efficient data generation using quantum measurements on qubits can solve problems in many areas, particularly finance. 
For example, quantum computers can be used for derivative pricing, risk modelling, and portfolio optimization \cite{herman2023quantum}.
In the study of the efficiency of Quantum computers in finance, there exists an important algorithm called Quantum amplitude estimation (QAE) algorithm \cite{suzuki2020amplitude}.

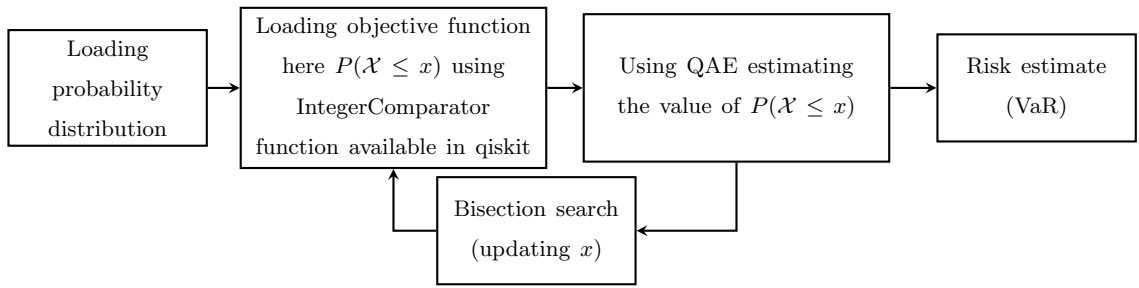
\begin{figure}[h]
    \tikzstyle{block1} = [draw, rectangle, 
    minimum height=5em, minimum width=3.8em,text width = 4cm,text centered]
    \tikzstyle{block2} = [draw, rectangle, 
    minimum height=1.5cm, minimum width=2.5cm,text width = 2.5cm,text centered]
    \tikzstyle{arrow} = [thick,->,>=stealth]
    \centering
    \begin{tikzpicture}[thick,scale= 2, every node/.style={scale=0.95}]
        \node (box1) [block2] {\fontsize{9}{2}\selectfont Loading probability distribution};
        \node (box2) [block1,xshift =4cm] {\fontsize{9}{2}\selectfont Loading objective function here   $P(\mathcal{X} \leq x)$ using IntegerComparator function available in qiskit};
        \node (box3) [block1,xshift =8.8cm] {\fontsize{9}{2}\selectfont Using QAE estimating the value of $P(\mathcal{X} \leq x)$};
        \node (box4) [block2,yshift = -2cm,xshift =6cm] {\fontsize{9}{2}\selectfont Bisection search (updating $x$)};
        \node (box5) [block2,xshift =13cm] {\fontsize{9}{2}\selectfont Risk estimate (VaR)};
        \draw [arrow] (box1)-- (box2);
        \draw [arrow] (box2)-- (box3);
        \draw [arrow] (box3) |- (box4);
        \draw [arrow] (box4) -| (box2);
        \draw [arrow] (box3) -- (box5);
    \end{tikzpicture}
      \caption{Flowchart representing all the steps involved in calculating VaR using a quantum computer.}
    \label{fig:flow_chart}
\end{figure}

The QAE algorithm promises quadratic speed-up over the classical Monte Carlo algorithm, which has applications in finance to price an option \cite{stamatopoulos2020option} or to calculate risk metrics like Value at Risk (VaR) and Conditional Value at Risk (CVaR) \cite{woerner2019quantum}. 
Fig. \ref{fig:flow_chart} shows the block diagram of the steps taken in calculating VaR with amplitude estimation.
First, we load the probability distribution of interest into the quantum computer and implement the objective function in the second step.
Then, the QAE block estimates the value of the objective function at each value of $x$ (for this, we use bisection search). 
The second, third and fourth steps are iterative and stop when the desired value of $x$ is reached, which is the required VaR value.
Hence, following the steps outlined in Fig. \ref{fig:flow_chart}, the QAE algorithm can be used to calculate the risk metric VaR, which can be extended to calculate CVaR \cite{woerner2019quantum}.
Apart from finance and risk management, the QAE algorithm finds application in all other fields where the Monte Carlo algorithm is used for calculation. 
However, the success of the QAE algorithm depends on the fact that there exists an algorithm which can be used to design a quantum circuit with as minimum circuit depth as possible to prepare a quantum register in a given target probability distribution.

The first step to using the QAE algorithm is to prepare a given quantum register in a probability distribution specific to the problem of interest, which we refer to as the loading problem. 
Suppose the probability distribution of interest comes under the family of log-concave probability distributions.
Then, we can use the well-known Grover-Rudolph (GR) \cite{grover2002creating} state preparation method to load the probability distribution. 
However, the GR state preparation method heavily depends on the Uniformly Controlled  Rotational $Y$ ($R_y$) gates.
Uniformly controlled rotational $R_y$ gates having $k$ control qubits can be implemented using $2^k$ CNOT gates and $2^k$ single qubit rotational  $ R_y$ gates \cite{mottonen2005transformation}.  
We can't use the GR state preparation method if the probability distribution does not come under the family of log-concave probability distributions.
Also, for such probability distribution, we would require an exponential number of gates to load the distribution into a quantum register \cite{PhysRevA.83.032302,vazquez2021efficient}.
A closer look into the GR state preparation method reveals that implementing the method using only two-qubit (CNOT) and one-qubit gates on current Noisy Intermediate-Scale Quantum computer (NISQ) hardware will also require an exponential number of gates. 
For example, we will require 16 four-qubit controlled $R_y$ gates to prepare a quantum superposition state of five qubits. 
Implementation of each four qubit control gate will require approximately $2^4 = 16$ CNOT gates and $2^4 = 16$ single qubit rotation gate. 
Hence, to implement the GR method on the current NISQ hardware, we need to reduce the circuit depth further in terms of CNOT and single-qubit gates. 
To calculate the circuit depth of the circuits presented in this article, we will consider a circuit library consisting of Controlled-Not (CX/CNOT) gates and one-qubit gates.

The current NISQ computers, based on superconducting qubits technology, are expected to scale up to 10k-100k qubits by 2026.
At the same time, the decoherence time for the qubits is going to be limited \cite{ibmwebsite}.  
This limitation inspires us to develop methods that reduce circuit depth by utilising more qubits. 
Hence, in this work, we have devised a scheme to reduce the circuit depth requirement for bell-shaped probability distributions like the normal distribution.
We have achieved this by adding an extra classical step that deconvolves the PMF.
Our approach draws inspiration from the principle of deconvolution of a discrete-time sequence into two discrete-time sequences.
In this manuscript, we introduce a classical pre-processing step of deconvolution of Probability Mass Function (PMF), which decreases the circuit depth at the cost of using more qubits and is compatible with any state preparation method. 
This method is more compatible with today's  NISQ computers because the NISQ architecture is envisaged to have more qubits but a lower operating time.
We want to emphasise the fact that we are not introducing any new state preparation method. 
Rather, we are introducing a classical pre-processing step that can give us a circuit depth reduction when combined with any state preparation method for loading a bell-shaped probability distribution. 

The structure of the presented manuscript is as follows.
In Section \ref{sec: Related_Works}, we discuss some existing methods for state preparation and point out why they are incompatible with the current NISQ architecture. 
Following this, in Section \ref{sec: Motivation}, we outline the motivation for exploring the concept of deconvolution in preparing a quantum register in a given bell-shaped probability distribution. 
Then, in  Section \ref{sec: trust_region}, we discuss two completely different approaches for deconvolving a given target probability mass function. 
One approach is to reformulate the problem as a constrained optimization problem and then make use of an optimization algorithm package available in Python to deconvolve the target PMF.
The second approach is to look at the deconvolution of PMF from the polynomial factorization lens. 
To validate our approach and claim in Section \ref{sec: expt_results}, we discuss some experiments that were carried out on the Qiskit quantum simulator and IBM quantum hardware. 
Particularly, we look at the circuit depth required to load the Gaussian and Laplace distribution using the GR state preparation method with and without including the classical pre-processing step of deconvolution.
This is followed by conclusions and discussion in Section \ref{sec: conclude}. 
\section{Related Work} \label{sec: Related_Works}
State preparation has been a long-known problem for practical quantum computing. 
A lot of different approaches have been proposed in this area. 
One of the earliest solutions to this problem was suggested by Grover \cite{grover2000synthesis} and is called the Black Box state preparation algorithm. 
The solution can prepare a given superposition of $N$ quantum states using $O(\sqrt{N})$ steps.
The circuit depth of each step in terms of one-qubit and two-qubit gates needs to be calculated. 
Based on the Toffoli gate ($3$ - qubit gate) count, the modified black box state preparation method proposed in \cite{sanders2019black} is an improvement over Grover's black box state preparation algorithm.
From an initial inspection, the circuit depth of the state preparation circuit constructed using the Black Box algorithm and modified black box algorithm is expected to be very large because of their dependence on repeated usage of the Quantum Amplitude Amplification (QAA) subroutine.
The QAA subroutine is used again in the QAE algorithm, which further increases circuit depth for practical use cases like the calculation of VaR described in Figure \ref{fig:flow_chart}.
 This makes the black box state preparation method incompatible with such use cases, at least for the current NISQ architecture. 
The quantum state preparation algorithm based on block encoding proposed in \cite{mcardle2022quantum}, unlike the black box state preparation algorithm, does not require an amplitude oracle.  
This leads to saving a considerable amount of quantum resources in terms of gates and qubits \cite{mcardle2022quantum}. 
The block encoding state preparation algorithm is compatible with early fault-tolerant quantum computers due to its dependence on a large number of Toffoli gates.
Therefore, it cannot be used on current quantum computers that belong to the NISQ era for quantum state preparation. 

Recently, there has been a lot of activity in the quantum computing community to solve the data loading problem using techniques from other fields like Tensor Networks \cite{melnikov2023quantum,gonzalez2024efficient}.
Quantum algorithms based on the Tensor network aim to make efficient use of Matrix Product State (MPS) representation of state vectors to tackle various problems in quantum computation like state preparation, simulation of large quantum circuits, etc.
Tensor Networks is a relatively new field with the potential to simulate quantum algorithms on classical computers efficiently, but further studies need to be done, and hence, the Tensor network approach falls beyond the scope of this study.
 A divide-and-conquer approach to the state preparation problem was proposed by Araujo \emph{et al.}, providing an exponential time advantage with a quantum circuit having poly-logarithmic circuit depth \cite{araujo2021divide}. 
 But the state preparation circuit designed using the algorithm scales in space (no. of qubits) as $O(N)$, where $N$ is the dimension of the state vector in which the quantum state should be prepared. 
 Another approach to solving the state preparation problem is using quantum Generative Adversarial Networks (qGAN) \cite{zoufal2019quantum}. 
The qGAN approach requires a circuit depth of $O\left(poly\left(\mathfrak{n}\right)\right)$ to prepare an $\mathfrak{n}$ qubit quantum register in a given probability distribution.
Due to the efficiency of this method in terms of circuit depth, the qGAN approach is very much compatible with quantum algorithms like QAE and HHL.
Even though it is efficient in terms of circuit depth, it requires a training step, making it a time-consuming approach. 
Additionally, as we scale the number of qubits $\mathfrak{n}$, there is a chance of encountering a barren plateau. 
Hence, further studies need to be carried out to determine the optimal quantum generator, discriminator structure, and training strategy. These shortcomings of the existing state preparation algorithm motivate us to design new algorithms that are more compatible with the NISQ architecture.

In derivative pricing and risk management, the most common and straightforward approach used for state preparation is the previously mentioned Grover Rudolph (GR) state preparation algorithm. 
The GR state preparation method, which was first introduced in \cite{grover2002creating}, since then, has undergone many modifications aimed at achieving an efficient implementation of the algorithm on a gate-based quantum computer.
Further circuit depth reduction for loading bell-shaped probability distribution can be achieved by combining the classical pre-processing step of deconvolution proposed in the manuscript with the GR state preparation method. 
The results ascertaining this claim are explained in Section \ref{sec: expt_results}.

\section{Notation}
\begin{enumerate}
    \item Random variables by $\mathcal{X}$, $\mathcal{Y}$, $\mathcal{Z}$
    \item Pauli gates by $X$, $Y$, $Z$.
    \item Probability mass functions by $\boldsymbol{P}, \boldsymbol{R}, \boldsymbol{q}$
    \item Number of qubits in a quantum register $\mathfrak{a}$, $\mathfrak{b}$, $\mathfrak{n}$.
    \item $\left \lceil  \hspace{1mm} \right \rceil$ represents the ceiling function.
    \item $\left \lfloor \hspace{1mm} \right \rfloor$ represent the floor function.
\end{enumerate}
\section{Hybrid classical quantum algorithm}\label{sec: Motivation} 
In this work, we try to solve the loading problem for bell-shaped probability distribution using the principle of deconvolution of discrete signals. 
Bell-shaped probability distributions are used to model the behaviour of random variables, which are used to calculate risk metrics like VaR and CVaR. 
Hence, we believe our work further facilitates the usage of quantum computers for risk management.  
As mentioned earlier, an important algorithm that depends on the efficient loading of an arbitrary probability distribution is the QAE algorithm, and it is used in the calculation of VaR.
QAE algorithm provides a quadratic speed-up compared to the Monte Carlo algorithm in simulating complex systems \cite{vazquez2021efficient}. 
We explain the QAE algorithm in more detail and stress its importance in \ref{sec: QAE_Algorithm}.
An efficient solution to the loading problem may expedite the adoption of the QAE algorithm. 

The complexity of creating an arbitrary quantum state is exponential in the number of qubits. 
However, there exists a trade-off between the number of qubits required for loading and time in terms of circuit depth. 
Our approach uses a hybrid scheme consisting of a classical pre-processing step followed by a quantum circuit. 
In this way, we attempt to integrate classical and quantum computers.
As proof of concept for our approach, we use the classical constraint optimization algorithm, namely the trust region method, to deconvolve the PMF and design a comparatively shallow quantum circuit for loading probability distribution at the cost of using more qubits.
The classical pre-processing step runs iteratively until the chosen cost function is minimized.
This cost-function-dependent outcome is used to design a quantum circuit that loads the required probability distribution into the quantum computer. 
In many cases, this circuit acts as a subroutine in quantum algorithms like QAE to calculate quantities of real-world importance. 
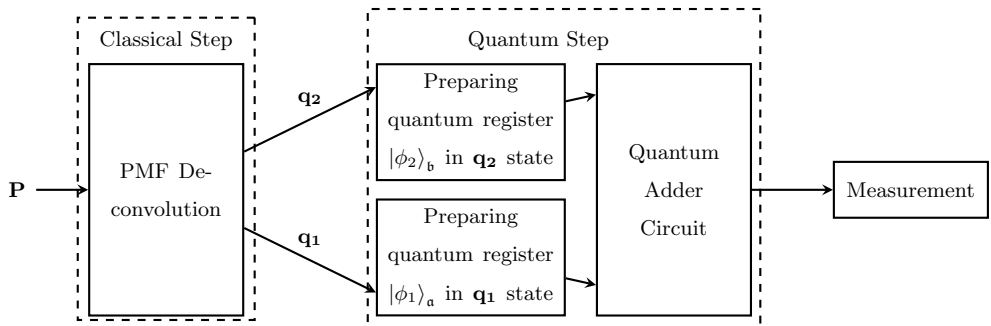
\begin{figure}[h]

 \tikzstyle{block1} = [draw, rectangle, 
     minimum height=9em, minimum width=4em,text width = 2cm,text centered]
     \tikzstyle{block2} = [draw, rectangle, 
     minimum height=2em, minimum width=5em,text width = 2.5cm,text centered]
     \tikzstyle{block3} = [draw, rectangle, 
     minimum height=2em, minimum width=3.2em,text width = 2cm,text centered]
     \tikzstyle{arrow} = [thick,->,>=stealth]
     \centering
     \begin{tikzpicture}[thick,scale=1.5, every node/.style={scale=0.9}]
        \node [xshift = -2.2cm](P) {\fontsize{9}{2}\selectfont $\bold{P}$}; 
         \node (box1) [block1] {\fontsize{9}{2}\selectfont PMF Deconvolution};
         \node (box2) [block2,yshift =-1.0cm,xshift =4.5cm] {\fontsize{9}{2}\selectfont Preparing quantum register $\ket{\phi_1}_{\mathfrak{a}}$ in $\bold{q_1}$ state };
         \node (box3) [block2,yshift = 1.0cm,xshift =4.5cm] {\fontsize{9}{2}\selectfont Preparing quantum register $\ket{\phi_2}_{\mathfrak{b}}$ in $\bold{q_2}$ state};
         \node (box5) [block3,xshift = 11cm] {\fontsize{9}{2}\selectfont Measurement};
         \draw [arrow] ([yshift = -1.6cm] box1)  -- node[yshift = 3mm]{\fontsize{9}{2}\selectfont $\bold{q_1}$} (box2);
         \draw [arrow] ([yshift = 1.6cm] box1)  -- node[yshift = 3mm]{\fontsize{9}{2}\selectfont $\bold{q_2}$} (box3);
         \node (box4) [block1,xshift = 7.5cm] {\fontsize{9}{2}\selectfont Quantum Adder Circuit};
         \draw[arrow] ([yshift = -1.6cm ]box2) -- (box4);
         \draw[arrow] ([yshift = 1.6cm ]box3) -- (box4);
         \node[above of = box4,xshift = -2cm,yshift = 1.2cm] (A) {\fontsize{9}{2}\selectfont Quantum Step};
         \node[fit= (A) (box3) (box4), dashed,draw,inner sep=0.40cm] (Box){};
         \draw [arrow] (P) -- (box1); 
         \node[above of = box1,yshift = 1.2cm] (B) {\fontsize{9}{2}\selectfont Classical Step};
        \node[fit= (box1) (B), dashed,draw,inner sep=0.27cm] (Box){};
         \draw[arrow] (box4) -- (box5);
     \end{tikzpicture}
    \caption{Flowchart representing all the steps involved. 
    Here, $\bold{P}$ (input) is the target probability distribution.}
    \label{fig: my_label}
\end{figure}

In Fig. \ref{fig: my_label}, we sketch an outline of the hybrid classical-quantum algorithm proposed in the manuscript. 
The classical step consists of deconvolving the target PMF into two smaller PMFs, sharing the task of generating a bell-shaped probability distribution among two quantum registers.
However, unlike depicted in Fig \ref{fig: my_label}, our approach can be extended to deconvolve the target PMF into more than two smaller PMFs, which will further reduce the circuit depth. 
The classical step is explained in detail in Section \ref{sec: deconvolution}, and the quantum step is presented in the following section.

\subsection{Preparation of quantum registers} \label{sec: Preparation of quantum registers}
For loading a given distribution into a quantum register having $\mathfrak{n}$ qubits, we have to discretize the probability distribution into $2^\mathfrak{n}$ regions. 
This will produce a Probability Mass function (PMF) having $2^\mathfrak{n}$ entries, and in terms of quantum computing, it will correspond to the square of the state vector of an $\mathfrak{n}$ qubit quantum system.  
In this article, we will concentrate on probability distributions that lie in the intersection of the bell-shaped probability distribution and the log-concave probability distribution families.
The log-concave probability distribution family consist of distributions that are efficiently integrable using existing classical integration techniques and hence can be discretized efficiently using a classical computer \cite{grover2002creating}.
Once we have discretized the probability distribution and obtained a state vector $\ket{\psi}_\mathfrak{n}$ of length  $2^\mathfrak{n}$, the next step is to prepare a quantum circuit that will transform the state $\ket{0}_\mathfrak{n}$ to $\ket{\psi}_\mathfrak{n}$. 
State preparation methods like the GR state preparation method, mentioned in the earlier sections, and those based on the GR state preparation method use multi-controlled $R_y$ gates, one after another, to load a given distribution into a $\mathfrak{n}$ qubit quantum system. 
The implementation of the multi-controlled $R_y$ gate on the current NISQ hardware requires us to decompose it using the gates mentioned in the circuit library.  
As per the GR state preparation method, these multi-controlled $R_y$ gates are applied sequentially, leading to an increase in the circuit depth.
To demonstrate this in \ref{app: GR_State_Preparation_Method}, we prepare the state of a two-qubit and three-qubit system using the GR state preparation method.
\begin{figure}[h]
    \centering
    \includegraphics[scale = 0.3]{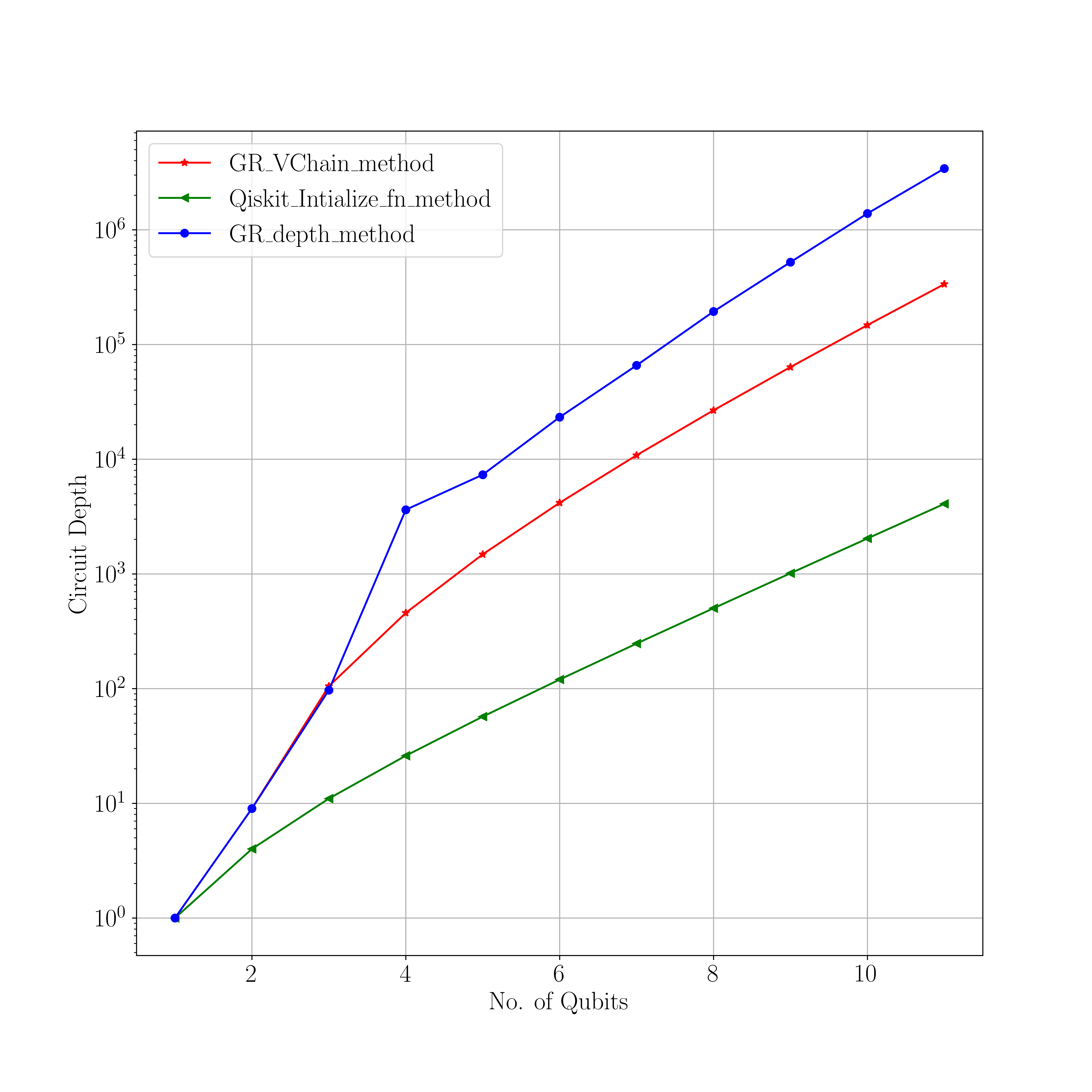}
    \caption{On the X-axis, we plot the number of qubits that we have to prepare in a given superposition state, and on the Y-axis, we plot the circuit depth required to prepare these qubits in a given superposition state. We use the Grover Rudolph state preparation method to prepare the state of the qubits in a given probability distribution.
    To implement the multi-controlled $R_y$ gate, we use the VChain method depicted in Fig. \ref{fig: vchain} in Appendix \ref{sec: VChain}.}
    \label{fig: GR_depth_vs_qubit_graph}
\end{figure}
Fig. \ref{fig: GR_depth_vs_qubit_graph} illustrates the scaling of circuit depth for different state preparation methods. 
For this, we plot a graph between the circuit depth of the circuit designed by using different state preparation methods and the number of qubits ($\mathfrak{n}$) in the quantum register $\ket{\psi}_\mathfrak{n}$.
The blue, green, and red plots represent the scaling of circuit depth concerning several qubits that are being prepared in a given target probability distribution state using different state preparation methods like the Grover Rudolph (GR) state preparation method, qiskit's built-in \texttt{initialize} function \cite{Intiliaze} and the Grover Rudolph method with VChain implementation, respectively.  
The difference between the GR and GR with VChain implementation is how we decompose the multi-controlled $R_y$ gate. 
In the normal GR implementation, we do not use ancilla qubits and decompose the multi-controlled $R_y$ gate using CNOT and various single qubit-quantum gates. 
We can also use another approach in which we use ancilla qubits to decompose the multi-controlled $R_y$ gate called the VChain method. 
The method is depicted in Fig. \ref{fig: vchain} and explained briefly in the \ref{sec: VChain}.
In the VChain approach, the Toffoli ($CCX$) gates are further decomposed into single qubit and CNOT gates for hardware implementation. 
But still, as expected from the MCMT documentation available on the Qiskit website\cite{Qiskitwebsite}, we see a circuit depth reduction. 

The state of a quantum register having a single qubit can be prepared using a $R_y$ gate. 
But, to prepare a superposition state of 5 qubits where we use all the $2^5$ computational basis states available in the $5$ qubit Hilbert space, we need a quantum circuit having a circuit depth in terms of CNOT and single-qubit gates in $O(2^5)$.
Similarly, for preparing a $13$ qubit system in a given distribution using the GR method, we need $O(2^{13})$ gates, which is huge for current NISQ computers.
Using Qiskit's in-built function called ``initialize" instead of the Grover Rudolph method results in a relatively reduced circuit depth. 
The Qiskit's in-built function \texttt{qiskit.extensions.Initialize()} is based on the method described in \cite{shende2005synthesis}. 
Apart from the method described in \cite{shende2005synthesis}, Qiskit's \texttt{qiskit.extensions.Initialize()} function also makes use of some circuit optimization techniques to cut down the circuit depth further. 
However, it's important to note that as the number of qubits increases, the depth of the circuit prepared by the Qiskit initialize function also scales up rapidly.
In terms of single-qubit and CNOT gates, the circuit depth of the GR method and Initialize function differ by a constant factor.

Now, let us briefly understand what deconvolution of PMF means.
To do so, we start by understanding the definition of convolution of probability mass functions.
Suppose we have two independent random variables, $\mathcal{X}$ and $\mathcal{Y}$, and we define another random variable, $\mathcal{Z}$, as the sum of the random variables, $\mathcal{X}$ and $\mathcal{Y}$.
Then, the probability distribution of the random variable $\mathcal{Z}$ is given by the convolution of the probability distributions of random variables $\mathcal{X}$ and $\mathcal{Y}$, respectively \cite{hajek2015random}. 
Deconvolution is the operation inverse to convolution. 
Given a probability distribution, can we split it into two probability distributions, each corresponding to an independent random variable?
In Section \ref{sec: trust_region}, we discuss an optimization approach to deconvolve a given PMF of length $N$ into two smaller PMFs of length $\lfloor \frac{N+1}{2} \rfloor$ and $\lceil \frac{N+1}{2} \rceil$. 
We load these two in parallel on two different quantum registers, then add the quantum data from these two registers and store the result of the adder in any of the registers using an additional qubit.
Loading these two distributions of smaller lengths requires a shallower quantum circuit compared to loading the target probability distribution directly without deconvolution using the quantum circuit designed by the GR state preparation method. 
\par
We take an $\mathfrak{a}$-qubit quantum register and a $\mathfrak{b}$-qubit quantum register, where $\mathfrak{a} = \left\lceil \log_2\left(\lfloor \frac{N+1}{2} \rfloor\right)\right\rceil$ and $\mathfrak{b} =\left\lceil \log_2\left(\lceil \frac{N+1}{2} \rceil \right)\right \rceil$.
Here, note that $\mathfrak{b} \geq \mathfrak{a}$ since $\frac{N+1}{2}$ is rounded off by the ceiling function.
In Section \ref{sec: expt_results}, where we discuss our results, we load PMFs of length $7$ and $15$.
Since the lengths are odd integers, the value of $\frac{N+1}{2}$ is an integer, and hence, for the odd length cases, we have $\mathfrak{a} = \mathfrak{b} = \left\lceil \log_2\left(\frac{N+1}{2}\right) \right\rceil$.
The quantum register comprising of $\mathfrak{a}$ qubits is initialized to the state $\ket{\phi_1}_\mathfrak{a}$, while the other quantum register, consisting of $\mathfrak{b}$ qubits, is set to the state $\ket{\phi_2}_\mathfrak{b}$.
Mathematically, the operation of preparing the quantum registers in the specified quantum states can be represented as
\begin{eqnarray}
    \ket{\phi_1}_\mathfrak{a} &= \sum_{i =0 }^{L(\boldsymbol{q_1})-1} \sqrt{q_{1_i}}\ket{i}_\mathfrak{a} = A\ket{0}_\mathfrak{a},\label{eq: expequation1}\\
    \ket{\phi_2}_\mathfrak{b} &= \sum_{j = 0}^{L(\boldsymbol{q_2})-1} \sqrt{q_{2_j}}\ket{j}_\mathfrak{b} = B\ket{0}_\mathfrak{b} \label{eq: expequation2},
\end{eqnarray} 
where  $L(\boldsymbol{q_1})$ and $L(\boldsymbol{q_2})$ represent the lengths of PMFs $\boldsymbol{q_1}$ and $\boldsymbol{q_2}$ respectively. 
In the above equations \ref{eq: expequation1} and \ref{eq: expequation2}, the vectors $\boldsymbol{q_1}$ and $\boldsymbol{q_2}$ are normalized.  
Using a quantum adder circuit represented by the unitary $U_A$, we add two quantum states defined in Equation \ref{eq: expequation1} and \ref{eq: expequation2}.
The construction of Quantum adder circuit $U_A$ is discussed in detail in the following Section \ref{sec: QuantumAdder}.
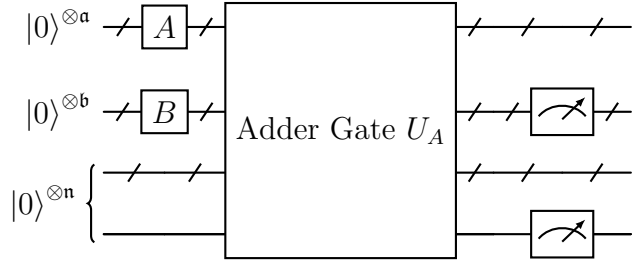
\begin{figure}[h]

    \centering
    \begin{tikzpicture}
    \node[scale = 1]{
    \begin{quantikz}
         \lstick{$\ket{0}^{\otimes \mathfrak{a}}$}  & \gate{A} \qwbundle[alternate]{}&\gate[4]{\text{Adder Gate $U_A$}}\qwbundle[alternate]{}&\qwbundle[alternate]{} &\qwbundle[alternate]{}&\qwbundle[alternate]{}\\
         \lstick{$\ket{0}^{\otimes \mathfrak{b}}$}  & \gate{B}\qwbundle[alternate]{}&\qwbundle[alternate]{}&\qwbundle[alternate]{} &\meter{}\qwbundle[alternate]{}&\qwbundle[alternate]{}\\
         \lstick[wires =2]{$\ket{0}^{\otimes \mathfrak{n}}$} & \qwbundle[alternate]{}& \qwbundle[alternate]{}&\qwbundle[alternate]{}&\qwbundle[alternate]{}&\qwbundle[alternate]{}\\
          & \qw& \qw&\qw &\meter{} &\qw
     \end{quantikz}
     };
     \end{tikzpicture}
 \caption{Quantum Circuit Loading probability distribution using the deconvolution method.
$A$ gate transforms the state $\ket{0}_\mathfrak{a}$ to state $\displaystyle \sum_{i=0}^{L(\boldsymbol{q_1})-1} \sqrt{q_{1_i}}\ket{i}_\mathfrak{a}$, 
similarly $B$ gate transforms the state $\ket{0}_\mathfrak{b}$ to state $\displaystyle \sum_{j=0}^{L(\boldsymbol{q_2})-1} \sqrt{q_{2_j}}\ket{j}_\mathfrak{b}$. $A$ and $B$ are constructed using the GR method.
The outcome of these two gates is passed into the adder circuit.  
 }
 \label{fig:quantumcircuit1}
\end{figure}

Fig \ref{fig:quantumcircuit1} depicts the quantum circuit design that implements the steps described so far on a gate-based quantum computer. 
Mathematically, we can break down the quantum circuit outlined in Fig \ref{fig:quantumcircuit1} and understand it further using the equations 
\begin{align}
     \left(A \otimes B \otimes \mathrm{I}_{\mathfrak{n}}\right)\ket{0}_\mathfrak{a}\ket{0}_\mathfrak{b}\ket{0}_\mathfrak{n} &= \left(\sum_{i=0}^{L(\boldsymbol{q_1})-1}\sqrt{q_{1_i}}\ket{i}_\mathfrak{a}\right)\otimes \left(\sum_{j=0}^{L(\boldsymbol{q_2})-1}\sqrt{q_{2_j}}\ket{j}_\mathfrak{b}\right)\ket{0}_\mathfrak{n}, 
\end{align}
\begin{align}
      U_A\left(\sum_{i=0}^{L(\boldsymbol{q_1})-1}\sum_{j=0}^{L(\boldsymbol{q_2})-1}\sqrt{q_{1_i}}\sqrt{q_{2_j}}\ket{i}_\mathfrak{a}\ket{j}_\mathfrak{b}\ket{0}_\mathfrak{n}\right) &= \sum_{i=0}^{L(\boldsymbol{q_1})-1}\sqrt{q_{1_i}}\ket{i}_\mathfrak{a}\sum_{h=0}^{L(\boldsymbol{q_1})+L(\boldsymbol{q_2})-2}\sqrt{k_h}\ket{h}_{\mathfrak{b}+1}\ket{0}_{\mathfrak{n}-1}. \label{eq: circuiteq}
\end{align}
The resulting quantum state after performing addition has a probability distribution obtained by the convolution of probability distribution $\boldsymbol{q_{1}}$ and $\boldsymbol{q_{2}}$, i.e., $\boldsymbol{k} = \boldsymbol{q_{1}} \ast \boldsymbol{q_{2}}$. 
The addition operation over qubits in equation \ref{eq: circuiteq}, denoted by $h = i + j$, is subsequently encoded into the quantum register $\ket{j}_\mathfrak{b}$.
Therefore, if we break down the probability distribution to be loaded into two smaller probability distributions, then all we need to do is load these two distributions using two separate GR state preparation circuits in parallel and add the two resulting states using the quantum adder circuit shown in Fig. \ref{fig: my_label}.

\subsection{Quantum adder}\label{sec: QuantumAdder}
\begin{figure}[h]
    \centering
    \begin{tikzpicture}
    \node[]{
    \begin{quantikz}
        \lstick[wires =2]{$\ket{\phi_1}_{\mathfrak{a}}$} &\qw &\ctrl{2} &\qw&\qw&\ctrl{2}&\qw &\qw&\qw&\qw&\qw&\qw\\
        \qw & \qw & \qw &\ctrl{2} &\ctrl{2} &\qw&\qw&\qw&\qw&\qw&\qw&\qw\\
        \lstick[wires =3]{$\ket{\phi_2}_{\mathfrak{b}}$} &\qw &\ctrl{3}&\qw&\qw&\targ{}&\qw&\qw&\qw&\qw&\meter{}&\qw\\
        \qw & \qw & \qw &\ctrl{3}&\targ{}&\ctrl{2}&\qw&\targ{}&\qw&\qw&\meter{}&\qw\\
        \qw & \qw & \qw & \qw &\qw & \qw &\ctrl{2}&\qw&\targ{}&\qw&\meter{}&\qw\\
        \lstick[wires =3]{$\ket{\mathrm{ancilla}}_{\mathfrak{b}}$}&\qw& \targ{} &\qw &\qw &\ctrl{1}&\qw &\ctrl{-2}&\qw&\qw&\qw&\qw\\
        \qw & \qw & \qw & \targ{}&\qw&\targ{}&\ctrl{1}&\qw&\ctrl{-2}&\qw &\qw&\qw\\
        \qw & \qw & \qw & \qw & \qw & \qw &\targ{}&\qw&\qw&\qw&\meter{}&\qw\\
    \end{quantikz}
    };
    \end{tikzpicture}
    \caption{Design for Quantum adder Circuit constructed using Algorithm \ref{alg: Quantumadder} for ${\mathfrak{n}} =2$. }
    \label{fig: adder_cir_n_2}
\end{figure}
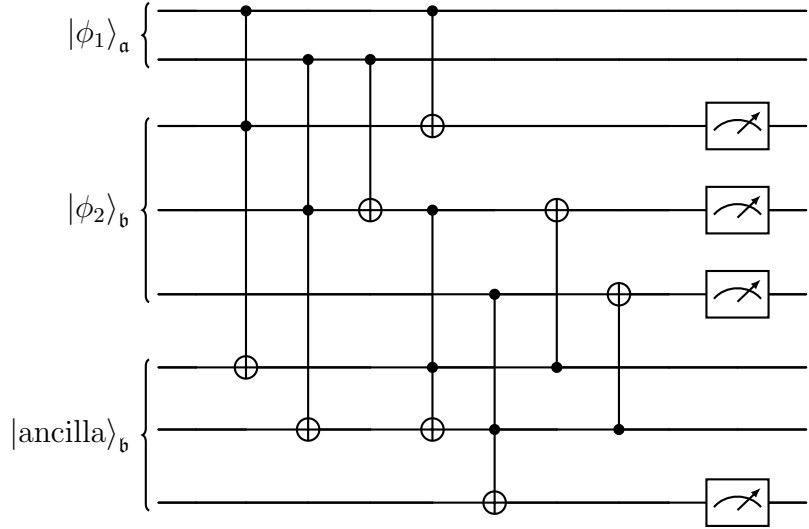

We present the design for the Quantum adder circuit shown in Fig.\ref{fig: my_label} and \ref{fig:quantumcircuit1} for adding two quantum registers $\ket{\mathrm{\phi_1}}_{\mathfrak{a}}$ and $\ket{\mathrm{\phi_2}}_{\mathfrak{b}}$, 
\begin{align} \label{eq: addercircuit_mat_eq}
   \ket{\mathrm{\phi_1}}_{\mathfrak{a}} \ket{\mathrm{\phi_2}}_{\mathfrak{b}} \ket{0}_{\mathfrak{b}} \xlongrightarrow{\text{Adder Circuit}} \ket{\mathrm{\phi_1}}_{\mathfrak{a}} \ket{\phi_1 \oplus \phi_2}_{\mathfrak{b}+1}\ket{\mathrm{garbage}}_{\mathfrak{b}-1},
\end{align}
where $\phi_1$ and $\phi_2$ refer to the quantum state of the quantum registers $\mathfrak{a}$ and $\mathfrak{b}$ in the computational basis state. 
The ``$\bigoplus$" in Equation \ref{eq: addercircuit_mat_eq} represents bit-wise modulo-2 addition.
The design for the quantum adder circuit used in this work is inspired from the VBE adder circuit described in \cite{vedral1996quantum}, but using this algorithm, we can add quantum registers of the same sizes.
This implies we can add quantum registers having $\mathfrak{a} = \mathfrak{b}$. 
In Algorithm \ref{alg: Quantumadder}, we modify the VBE approach a little bit so that we can work with unequal quantum registers $(\mathfrak{a} \neq \mathfrak{b})$. Additionally, an important point to note is that algorithm \ref{alg: Quantumadder} uses only CNOT and Toffoli gates for the construction of the quantum adder circuit.
In  Fig. \ref{fig: adder_cir_n_2} using Algorithm \ref{alg: Quantumadder}, we design a quantum circuit for adding quantum register having $\mathfrak{a} =2$ and $\mathfrak{b} =3$ qubits.
In this design, we note that we do not reset the ancilla qubits used by the adder circuit.
To reset the ancilla qubit, we need extra Toffoli gates, which will lead to an increase in the circuit depth. 
We avoid doing this since the main goal of the manuscript is to implement the state preparation approach with minimum circuit depth. Hence, we avoid adding gates that do not contribute to the computation. 
Equation \ref{eq: addercircuit_mat_eq} gives the mathematical representation for the action of the quantum adder circuit.
\begin{algorithm}
    \caption{Algorithm for designing Quantum adder Circuit to add two unequal Quantum registers} \label{alg: Quantumadder}
    \textbf{Input: } $\mathfrak{a}$ qubit quantum register prepared in $\ket{\phi_1}_\mathfrak{a}$ quantum state and $\mathfrak{b}$ qubit quantum register prepared in $\ket{\phi_1}_\mathfrak{b}$ quantum state. 1 extra qubit $\ket{0}$, which is included with the ancilla qubit. 
    \begin{algorithmic}
        \Require ${\mathfrak{b}}$ ancilla qubits
        \State $i \gets 0$
        \While{$i \neq \mathfrak{a}$}
            \State $\mathrm{CCX gate} $( Control qubit =  ($\ket{\mathrm{\phi_1}}_{\mathfrak{a}}[i], \ket{\mathrm{\phi_2}}_{\mathfrak{b}}[i]$),Target = $\ket{\mathrm{ancilla}}_{\mathfrak{b}}[i]$)
            \State $i \gets i + 1$
        \EndWhile
        \State $i \gets 0$
        \While{$i \neq {\mathfrak{a}}$}
            \State $\mathrm{CX gate} $( Control qubit =  $\ket{\mathrm{\phi_1}}_{\mathfrak{a}}[\mathfrak{a}-i-1]$,Target =  $\ket{\mathrm{\phi_2}}_{\mathfrak{b}}[\mathfrak{a}-i-1]$)
            \State $i \gets i + 1$
        \EndWhile
        \State $i \gets 1$
        \While{$i \neq \mathfrak{b}$}
            \State $\mathrm{CCX gate} $( Control qubit =  ($\ket{\mathrm{\phi_2}}_{\mathfrak{b}}[i]$,$\ket{\mathrm{ancilla}}_{\mathfrak{b}}$[i-1]),Target =  $\ket{\mathrm{ancilla}}_{\mathfrak{b}}[i]$)
            \State $i \gets i + 1$
        \EndWhile
        \State $i \gets 0$
        \While{$i \neq {\mathfrak{b}}$}
            \State $\mathrm{CX gate} $( Control qubit =  $\ket{\mathrm{ancilla}}_{\mathfrak{b}}[i]$,Target =  $\ket{\mathrm{\phi_2}}_{\mathfrak{b}}[i+1]$)
            \State $i \gets i + 1$
        \EndWhile
    \end{algorithmic}
\end{algorithm}
\par
\par
The circuit depth for implementing the GR state preparation method without using ancilla qubits and using only CNOT and single-qubit gates is given by  $\mathcal{O}(n2^{\mathfrak{n}})$ \cite{sun2023asymptotically}. 
At the same time, the circuit depth of the adder circuit scales linearly with the number of qubits \cite{vedral1996quantum}.
So, instead of using the quantum circuit designed by  GR state preparation method to prepare a state in $\ket{\psi}_{{\mathfrak{n}}+1}$ directly, we make use of two shallower quantum circuit designed by GR state preparation method in parallel and prepare two different quantum states $\ket{\phi_1}_{\mathfrak{a}}$ and $\ket{\phi_2}_{\mathfrak{b}}$.
Then, we pass these two states through the quantum adder circuit to achieve the state $\ket{\psi}_{{\mathfrak{n}}+1}$.
We note in the current context from \cite{hajek2015random} that addition in basis space is convolution in probability space, i.e., when we perform an addition operation on two quantum states, the probability amplitudes corresponding to each basis state of the two quantum states will convolve with each other.
Therefore, we need only $\mathcal{O}((\mathfrak{n}-1)2^{{\mathfrak{n}}-1})$ gates for the GR part and $\mathcal{O}({\mathfrak{n}})$ gates for the adder circuit. 
In the Algorithm \ref{alg: Quantumadder}, the notation $\ket{\mathrm{\phi_1}}_{\mathfrak{a}} [i]$ refers to the $i^{\mathrm{th}}$ qubit of the quantum register $\ket{\mathrm{\phi_1}}_{\mathfrak{a}}$.
In the following section, we will discuss two approaches to deconvolution, one in which we use an optimization technique called trust region and the second in which we use polynomial factorization.

\section{Deconvolution of Target PMF}\label{sec: deconvolution}
We present the classical part of our hybrid algorithm, which employs classical techniques to accelerate the functioning of quantum circuits.
In this section, we explore two approaches to deconvolve a target PMF that we envisage to generate as the end goal of the quantum hardware. 
In the first approach, we reformulate the task of deconvolution of a given PMF as a constrained optimisation problem and use the trust region algorithm to find the optimum solution. 
This approach also serves the purpose of verifying the statement that the addition of quantum states represented in computational basis states results in the convolution of the probability amplitudes of the quantum states. 
Another, more efficient approach to perform deconvolution involves polynomial factorization.

\subsection{Deconvolution using trust region Method} \label{sec: trust_region}
In this section, we describe how to perform deconvolution of a given PMF using the trust region method \cite{gould1999solving}. 
The deconvolution of a given target probability distribution can be reformulated as a constraint optimisation problem and hence can be mathematically formulated as : 
\begin{align}
    \begin{aligned}
        \min_{\boldsymbol{q_1, q_2}} f(\boldsymbol{q_1},\boldsymbol{q_2}) = JS(\boldsymbol{P}||\boldsymbol{Q})  \hspace{1mm} \text{where} \hspace{1mm} \boldsymbol{Q} = \boldsymbol{q_1} \ast \boldsymbol{q_2},
    \end{aligned}
    \label{eq: q_1_q_2_equation}
\end{align}
where, $\boldsymbol{q_1}$ and $\boldsymbol{q_2}$ represents normalized vectors and $\ast$ represents convolution operator acting on $\boldsymbol{q_1}$ and $\boldsymbol{q_2}$  \cite{oppenheim1999discrete}. 
Also, $\boldsymbol{q_1}$ and $\boldsymbol{q_2}$ vectors in Equation \ref{eq: q_1_q_2_equation} corresponds to a PMF and hence $0<q_{1_i}<1,0<q_{2_j}<1$ for all values of $i = 0,1,2,\hdots,L\left(\boldsymbol{q_1}\right)$ and $j = 0,1,2,\hdots,L\left(\boldsymbol{q_2}\right)$.
$\boldsymbol{Q}$ is a PMF of length $L(\boldsymbol{q_1})+L(\boldsymbol{q_2})-1$.
$\boldsymbol{P}$ represents the target PMF that we require to deconvolve and hence is also a vector of length $L(\boldsymbol{q_1})+L(\boldsymbol{q_2})-1$. 
The function $JS(\boldsymbol{P}||\boldsymbol{Q})$ is called the Jensen Shannon distance and is given by
\begin{align}
    JS(\boldsymbol{P}||\boldsymbol{Q}) = DS(\boldsymbol{P}||\boldsymbol{R}) + DS(\boldsymbol{Q}||\boldsymbol{R}),
\end{align}
where
\begin{align} \label{eq: Rdef}
     \boldsymbol{R} = \frac{1}{2}\left(\boldsymbol{P} + \boldsymbol{Q} \right) .
\end{align}
Further, $DS(\boldsymbol{P}||\boldsymbol{R})$ can be defined as 
\begin{align} \label{eq: DSformula}
        DS(\boldsymbol{P}||\boldsymbol{R}) = \sum_{i=0}^{L(\boldsymbol{q_1})+L(\boldsymbol{q_2})-2} P_i\log\frac{P_i}{R_i},
\end{align}
where $P_i$ refers to the $i^{\mathrm{th}}$ element of the PMF $\boldsymbol{P}$ and $R_i$ refers to the $i^{\mathrm{th}}$ element of the PMF $\boldsymbol{R}$ constructed using equation \ref{eq: Rdef}.
We use the trust region method to find the optimum $\boldsymbol{q_1}$ and $\boldsymbol{q_2}$ such that it minimises the Jensen–Shannon distance function $f(\boldsymbol{q_1},\boldsymbol{q_2})$. 
We chose the Jensen-Shannon function as the cost function due to its symmetric nature. 
We calculate the gradient of the Jensen–Shannon distance with respect to both probability vectors $\boldsymbol{q_1}$ and $\boldsymbol{q_2}$. 
Each vector has $L(\boldsymbol{q_1})$ and $L(\boldsymbol{q_2})$ independent variables, respectively, as they both represent a probability mass function (PMF) with the number of elements equal to $L(\boldsymbol{q_1})$ and $L(\boldsymbol{q_2})$ respectively. 
Since they are PMFs, they have to satisfy the constraint $\displaystyle \sum_{i=0}^{L(\boldsymbol{q_1})-1} q_{1_i} = 1$ and $\displaystyle \sum_{i=0}^{L(\boldsymbol{q_2})-1} q_{2_i} =1$, which reduces the number of independent variables from $L(\boldsymbol{q_1})$ to $L(\boldsymbol{q_1})-1$ and $L(\boldsymbol{q_2})$ to $L(\boldsymbol{q_2})-1$ respectively. 
We pass the Jensen–Shannon distance function, the gradient, the Hessian matrix and the method we intend to use to the Python package \texttt{scipy.optimize.minimize}. 
We explain the calculation of the Gradient and Hessian functions in detail in \ref{sec: Gradient_Hessian_fn_calculation}.
We use the standard Trust region optimizer from \texttt{scipy} \cite{Scipy_optimize} to find the minimum of the Jensen–Shannon distance function. 
We explain the sequence in which we perform these steps in Algorithm \ref{alg: built_deconv_optimize_algo}.
\begin{algorithm}
\caption{Algorithm for deconvolution of PMF }
\label{alg: built_deconv_optimize_algo}
\begin{algorithmic}
    \State Input: $ P $ 
    \State variable declare $\boldsymbol{q_1} = \mathrm{Random}\text{ }\mathrm{Guess}$.
    \State variable declare $\boldsymbol{q_2} = \mathrm{Random}\text{ }\mathrm{Guess}$.
    \State Calculate the Gradient using \ref{eq : gradientequation}.  
    \State Calculate the Hessian Matrix using the equations from \ref{eq: hes1} to \ref{eq:hes2}.
    \State Pass the cost function, gradient vector, and Hessen matrix to the \texttt{scipy.optimize} package of Python. 
    \State Terminate after 1000 iterations.
\end{algorithmic}
\end{algorithm}

Deconvolution through optimization faces all the drawbacks faced by an optimization algorithm. 
As we increase the length of the target probability mass function, the no. of variables over which we intend to optimize also increases simultaneously. 
This increases the time taken by the optimization algorithm to converge to the global minimum and hence limits us from using the deconvolution approach for preparing a quantum register having a large number of qubits.
However, an alternative approach exists where we deconvolve a given target probability distribution by employing polynomial factorization.
\subsection{Deconvolution Using Polynomial Factorization}

We know that any given PMF can also be represented using a probability-generating function (PGF) \cite{hajek2015random}.
The PGF for a discrete distribution corresponds to a polynomial function. 
We can calculate the PGF represented by $f(x)$ corresponding to a discrete probability distribution $\boldsymbol{P}$ (PMF) using the equation 
\begin{align}
    f(x) = \sum_{i=0}^{L(\boldsymbol{P})-1}P_i x^i.
    \label{eq: pgf_calculation}
\end{align}
Equation \ref{eq: pgf_calculation} implies that if we have PMF of length $2^\mathfrak{n}$, then there exists a polynomial function of degree $2^{\mathfrak{n}}-1$ representing the PGF of the given PMF.
It is also possible to calculate the PMF from the PGF of a distribution using the equation 
\begin{align}
    P_i = \frac{\partial^i f(x)}{\partial^i x}\Bigg|_{x=0}.
\end{align}
In this context, we can reformulate the deconvolution of probability distribution as follows. 
Let $f(x)$ be the polynomial function representing the PGF of a given probability distribution. 
We need to factorize the function $f(x)$ into polynomials that have non-negative coefficients. Specifically, we need to find polynomials $\Tilde{f_i}(x)$ with non-negative coefficients such that $\mathrm{deg}(\Tilde{f}_i(x))<2^{\mathfrak{n}-1}$ for all $i=1,\dots,K$ and \begin{align}
    f(x)=\prod_{i=1}^{K}\Tilde{f}_i(x)
\end{align}.
Additionally, the sum of the degrees of all the factor polynomials must add up to $2^{\mathfrak{n}}-1$, i.e.,
 \begin{align}
      \sum_{i=1}^{K} \mathrm{deg} \left(\Tilde{f}_i(x)\right) = 2^{\mathfrak{n}}-1.
\end{align}
Since polynomial $f(x)$ have all positive coefficients, then according to Theorem 3 in \cite{pos_coeff}, $f(x)$ does not have a root in $\mathbb{R}^+$ including zero.
This implies that the linear factors of the polynomial will be of the form $(x+c)$ where $c$ is a real positive constant, i.e., polynomial $f(x)$ will have linear factors with positive coefficients alone.
It will also have quadratic polynomial factors of the form $x^2-bx+c$ and $x^2+bx+c$. 
Now, our task is to get rid of the quadratic polynomial of the form $x^2-bx+c$ by multiplying it with the appropriate $x^2+bx+c$ and linear factor polynomial to form higher order polynomials having positive coefficients.

We have developed an algorithm to factorize the target polynomial into polynomials having non-negative coefficients based on calculating all the roots of the target polynomial.
The algorithm also takes into account that there can be multiple possible combinations for multiplying polynomials with all positive coefficients that produce the target polynomial.
Therefore, the factorization of the target polynomial into polynomials of smaller degrees having all positive coefficients is not unique. 
\begin{algorithm}
\caption{Algorithm for deconvolution using polynomial factorization}
    \begin{algorithmic}[1]
        \State Input: Target Polynomial $f(x)$.
        \Procedure{FindRoots}{$\text{coefficients of the target polynomial}f(x)$}
        \State Use the \texttt{numpy.polyroot} function to find all the roots of the polynomial $f(x)$ given by the coefficients.
        \State \textbf{return} Roots 
        \EndProcedure
        \Procedure{GroupRoots}{\text{Roots}}
        \State Group the roots into three: Complex roots with real part positive, Complex roots with real part negative and real roots.
        \State \textbf{return} Comp\_root\_real\_pos,Comp\_root\_real\_neg,real\_root
        \EndProcedure
        \State Arrange the Complex root with a positive real part in ascending order with respect to the real term and call it $\mathrm{basket}_1$. 
        \State Merge the group of Complex roots with the negative real part and the group of real roots and call it $\mathrm{basket}_2$.
        \State Randomly shuffle the elements of the $\mathrm{basket}_2$.
        \While{$\text{length}(\mathrm{basket}_1) \geq 0$}
            \State \text{temp\_array} = $\mathrm{basket}_1$[0]
            \State temp\_boolean = \textbf{True}
            \While{temp\_boolean}
                \State random\_index = generate random integer between 0 and length($\mathrm{basket}_2$).
                \State temp\_array = temp\_array + $\mathrm{basket}_2$[random\_index]
                \State poly\_coeff = \texttt{numpy.polynomial.polyfromroots}(temp\_array)
                \State Delete the element $\mathrm{basket}_2$[random\_index] from the list $\mathrm{basket}_2$ 
                \If{if any(c.real $\leq$ 0 for c in poly\_coeff)}
                    \State temp\_boolean = \textbf{False}
                  \Else
                    \State temp\_boolean = \textbf{True}
                  \EndIf
            \EndWhile
            \State Delete the first element from  $\mathrm{basket}_1$.
            \State $\mathrm{basket}_2$.append(temp\_array)   
        \EndWhile 
        \State Create an empty array variable and name it list\_of\_poly\_factors = [ ]. 
        This list holds the factors of the target polynomial.
        \For{\texttt{i in $\mathrm{basket}_2$}}
            \State temp = \texttt{numpy.polynomial.polyfromroots}(i)
            \State list\_of\_poly\_factors.append(temp/sum(temp))
        \EndFor
    \end{algorithmic}
    \label{alg: Poly_deconv}
\end{algorithm}
The algorithm for deconvolution using polynomials is given in Algorithm \ref{alg: Poly_deconv}. 
The Python function \texttt{FindRoots} defined in the  Algorithm \ref{alg: Poly_deconv} returns a 2D array of roots where the conjugate pair complex roots are stored as a 1D array.
In Algorithm \ref{alg: Poly_deconv}, $basket_1$ and $basket_2$ are also 2D arrays of roots.
We randomly shuffle the 1D set of roots stored in the $basket_2$ and not the individual roots inside the 1D array. 
The \texttt{numpy.polyroot} python function used in Algorithm \ref{alg: Poly_deconv} takes the coefficients of a polynomial as input and returns all the roots of the entered polynomial $f(x)$ \cite{Numpy_poly_roots}.
Similarly, the Python function \texttt{numpy.polyfromroots} take roots as input and return the coefficients of a monic polynomial \cite{Numpy_poly_from_roots}. 
In \ref{sec: appendix_2}, we explain this algorithm in detail using an example.  
In the following section, we will discuss the results that we obtained by deconvolving the given PMF using the optimization technique discussed in Section \ref{sec: trust_region}. 
However, using the approach discussed in this section, we can speed up the classical process of deconvolution and get more accurate results. 
Deconvolution performed by this approach is more accurate since algorithms based on an optimization approach sometimes give solutions up to an error bound. 
The optimization approach is efficient when the parameters we optimize over are less in number, and the cost function is smooth without any local minimum. 
When the number of parameters is too large, the algorithm based on the optimization technique terminates when it reaches a specific iteration limit, and the algorithm outputs an approximate result for deconvolution. 
\begin{figure}[h]
    \centering
    \includegraphics[height = 0.6\linewidth, width = \linewidth]{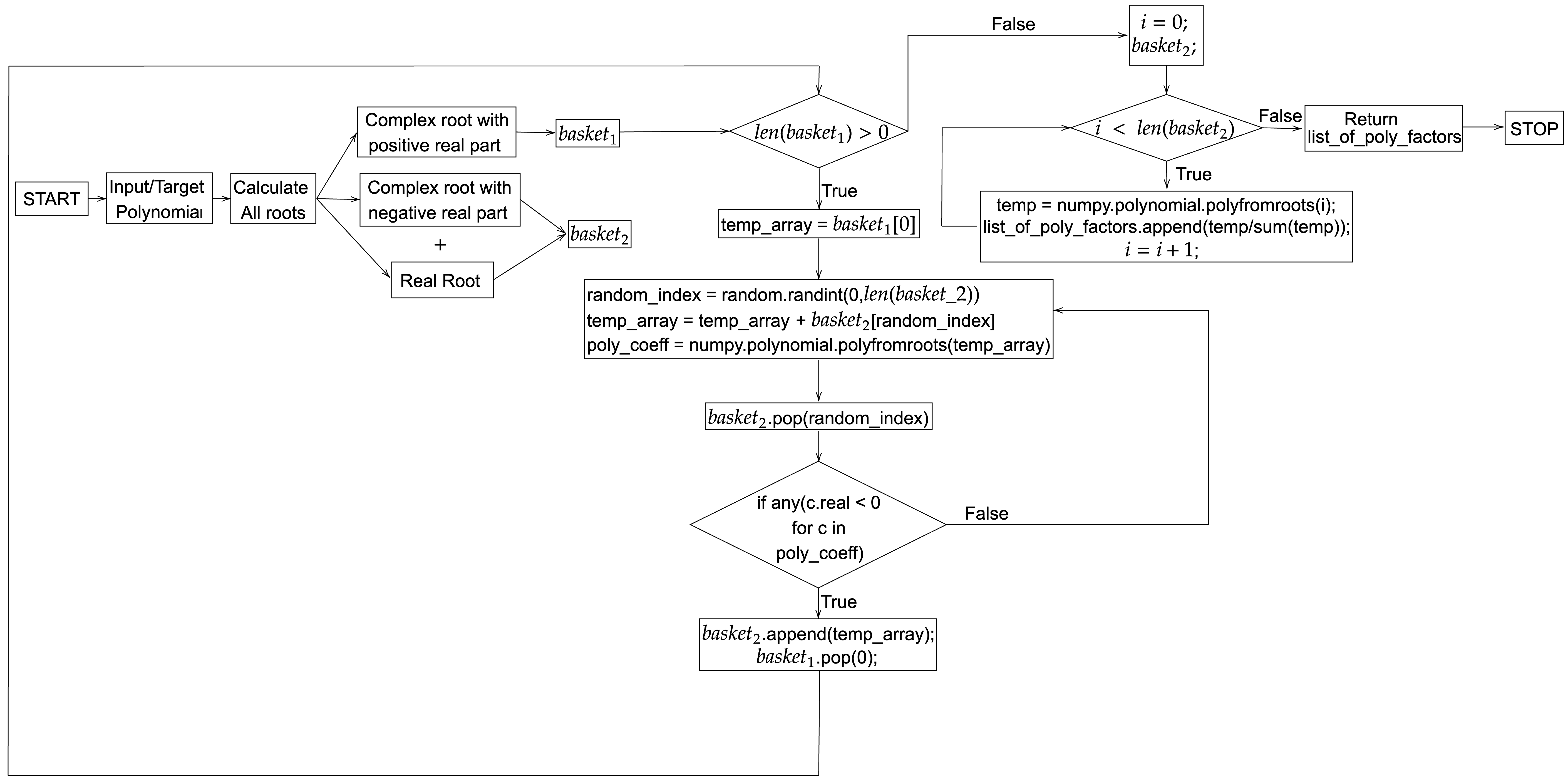}
    \caption{Flowchart for Algorithm \ref{alg: Poly_deconv}. }
    \label{fig:Flowchart_Deconvl_Algorithm}
\end{figure}

\section{Discussion of the experiments and results}\label{sec: expt_results}

\begin{figure}
    \centering
    \begin{subfigure}{0.475\textwidth}  
        \centering
        \includegraphics[scale=0.22]{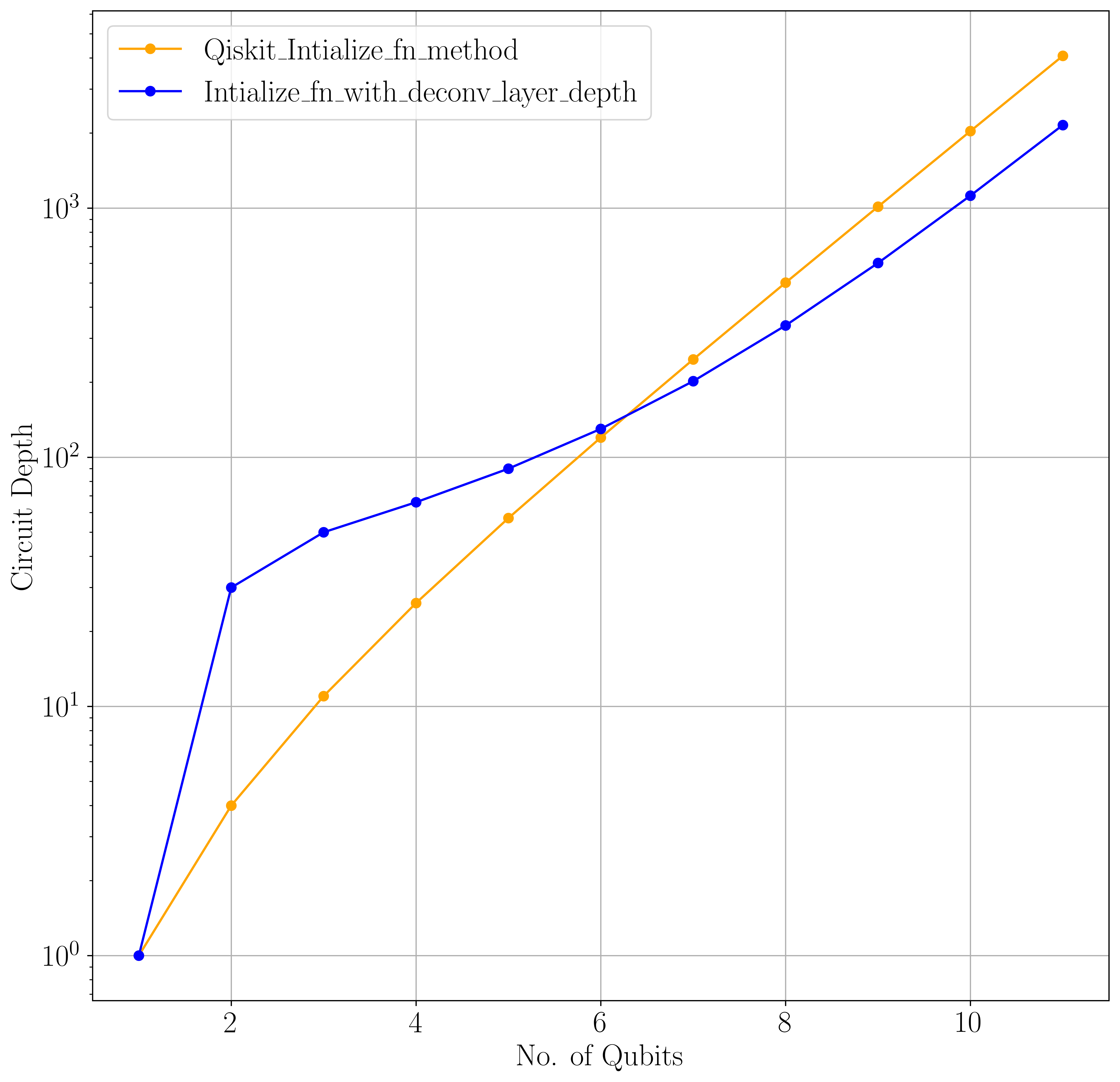}
        \caption{Qiskit Initialize function}
        \label{fig: Log_Qiskit_Intialize_fn_with_and_without_deconvolution}
    \end{subfigure}
    \hfill
        \begin{subfigure}{0.475\textwidth}  
        \centering
        \includegraphics[scale=0.22]{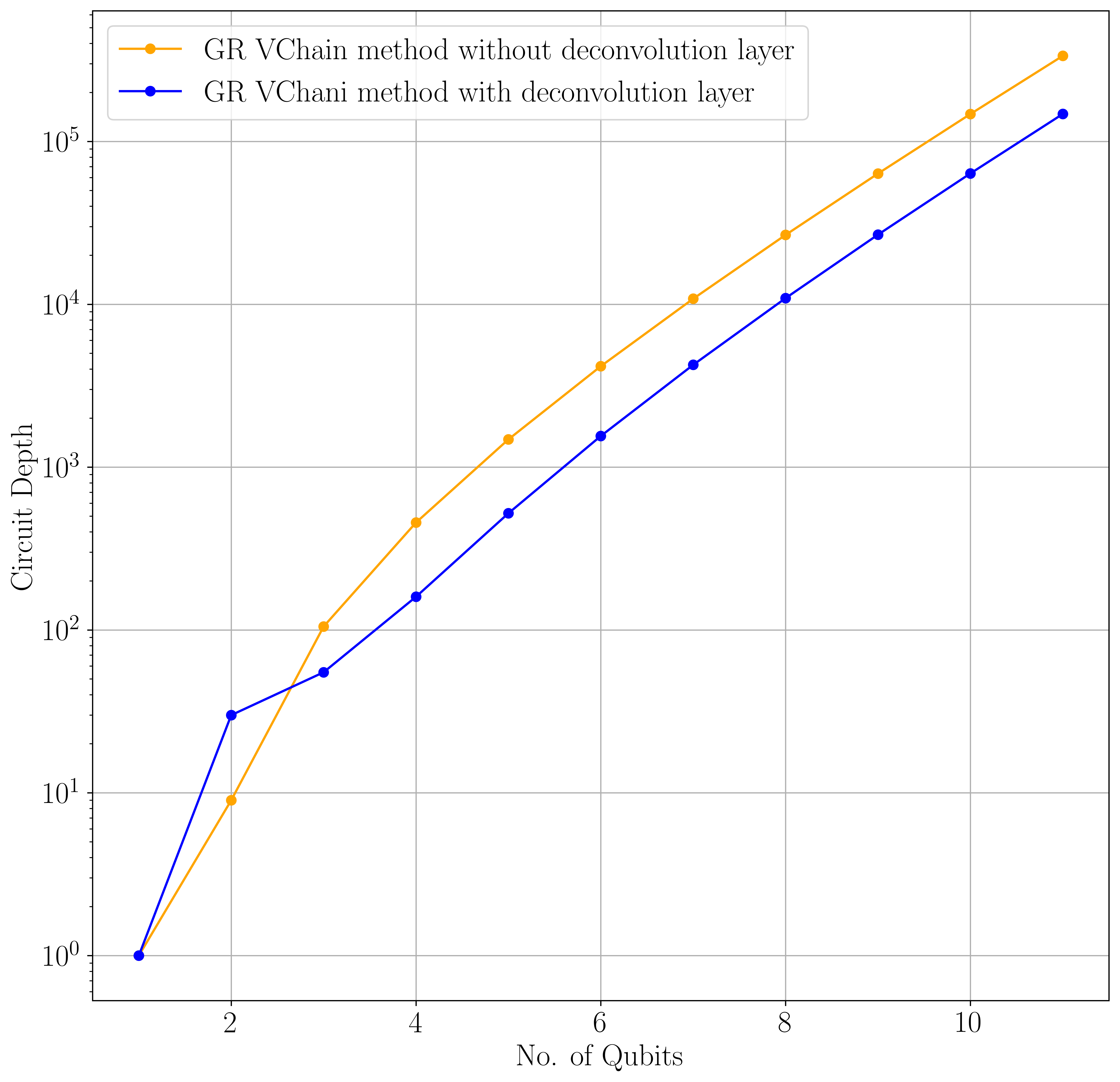}
        \caption{GR State preparation with VChain}
        \label{fig: Log_GR_with_without_deconvolution}
    \end{subfigure}
     \hfill
        \begin{subfigure}{0.475\textwidth}  
        \centering
        \includegraphics[scale=0.22]{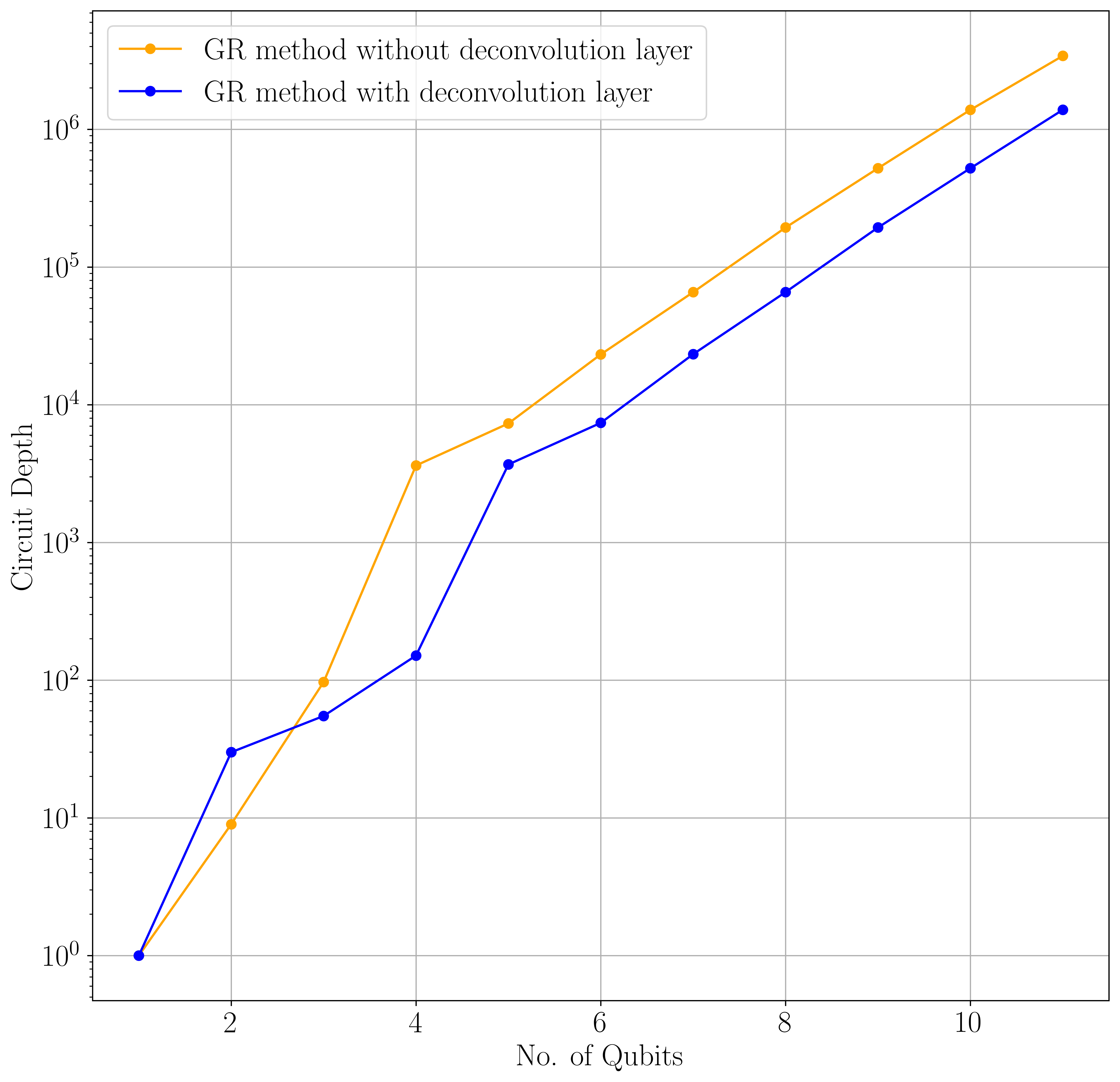}
        \caption{GR State preparation }
        \label{fig: Log_GR_VChain_with_without_deconvolution}
    \end{subfigure}
    \caption{We plot the Log of circuit depth Vs the number of qubit graphs for different state preparation methods. In Fig. (a), we use the Qiskit initialize function to prepare the qubits in a given state, (b) we use the GR state preparation method with VChain implementation for state preparation, and (c) we use the GR state preparation method.
    In the above figures, the orange line represents state preparation using the state preparation method without the deconvolution step, and the blue line represents state preparation using state preparation methods with an extra classical step of deconvolution. }
    \label{fig: log_deconvolution_advantage_graph}
\end{figure}

An extra classical deconvolution layer can reduce the circuit depth at the cost of using more ancillary qubits. 
To demonstrate this, in Fig. \ref{fig: log_deconvolution_advantage_graph}, we plot circuit depth versus the number of qubits that will be prepared in a given target probability distribution state with and without using the deconvolution layer. 
From Fig. \ref{fig: log_deconvolution_advantage_graph} we can infer that the state preparation method with the deconvolution layer performs better in terms of the circuit depth than the one without it. 
Hence, if the deconvolution layer is integrated with the GR state preparation method, then the combined state preparation method scales as $O((\mathfrak{n}-1)2^{\mathfrak{n}-1}+ \mathfrak{n})$, and therefore approximately halves the circuit depth.
In Fig. \ref{fig: Log_Qiskit_Intialize_fn_with_and_without_deconvolution}, \ref{fig: Log_GR_with_without_deconvolution} and \ref{fig: Log_GR_VChain_with_without_deconvolution} we plot the circuit depth scaling of Qiskit Initialize function, GR State preparation method and GR state preparation method with VChain implementation Vs the number of qubits whose state is being prepared in a target probability distribution with and without the deconvolution layer. 
As mentioned earlier in Section \ref{sec: Preparation of quantum registers}, the difference between the GR state preparation method and the GR state preparation method with VChain implementation is in how we implement the multi-controlled $R_y$ gate.
In the first case, we let Qiskit break down the multi-qubit gates into single-qubit and two-qubit (CNOT) gates without using any ancilla qubits. 
In the next case, we explicitly use the feature of multi-control multi-target V-Chains (MCMTVChain) in Qiskit to implement the multi-controlled $R_y$-gates as shown in Fig. \ref{fig: vchain}.
Here, we use ancilla qubits in a V-Chain structure to create a multi-controlled single target $R_y(\theta)$-gate.
\par
To test and compare the different state preparation methods with and without the deconvolution layer, we load Gaussian and Laplacian distributions on a QASM simulator and real IBM quantum hardware. 
We also calculate a metric called Quantum Circuit Volume (QCV) \cite{park2023quantum} for each implementation. 
The mathematical definition of QCV is given by, 
\begin{align}
    \mathrm{QCV}(C) = s(C) \times d(C).
    \label{eq: QCV_eq}
\end{align}
In equation \ref{eq: QCV_eq}, $C$ denotes the quantum algorithm for which we are implementing the quantum circuit, $s(C)$ represents the number of qubits we used for this implementation and $d(C)$ represents the depth of the quantum circuit implementation. 
The metric QCV is used to quantify the amount of quantum resources used to implement a quantum algorithm on a quantum computer.  
To test the method's performance on quantum hardware, real and simulated, we measured the quantum circuit 2048 times (shots).
Then, the JS distance between the empirical PMF and the input PMF is measured.
In section \ref{sec: trust_region}, we use the JS distance function as a cost function, which is minimized to deconvolve a given target probability distribution.
But unlike in section \ref{sec: trust_region}, here, JS distance is used as a metric to quantify the differences between the measured and target PMF. 
The circuits are run on IBMQ's noiseless QASM simulator and on \textit{ibmq\_kolkata}, which is one of the IBM Falcon Processors and has 27 qubits. 
The measured Quantum Volume of \textit{ibmq\_kolkata} processor is $128=2^7$. 
One important point to note is that Quantum Volume is different from Quantum Circuit Volume. 
Quantum volume is a metric that is used to compare different NISQ devices. 
The largest random circuit of equal depth and width that can be successfully implemented on a quantum computer is termed quantum volume \cite{quantumvolume}. 

\begin{figure}
  \centering
  \begin{subfigure}[b]{0.475\linewidth}
    \centering
    \includegraphics[width=\linewidth]{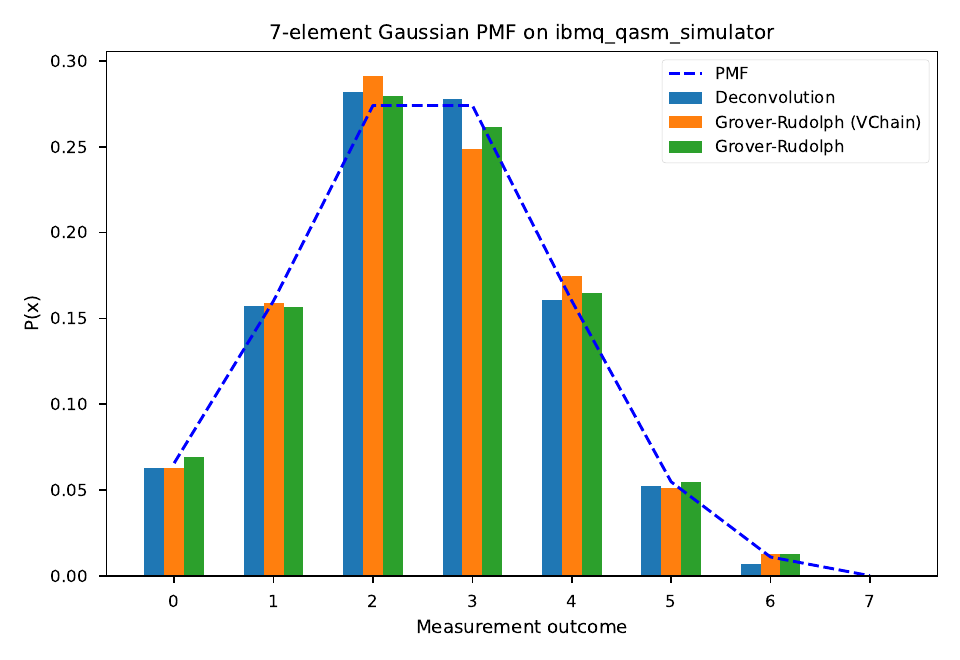} 
    \caption{7-element Gaussian PMF generated using IBMQ QASM simulator (Table \ref{tab:7-gaussian-sim}).} 
    \label{fig:7-gaussian-sim} 
  \end{subfigure}
  \hfill
  \begin{subfigure}[b]{0.475\linewidth}
    \centering
    \includegraphics[width=\linewidth]{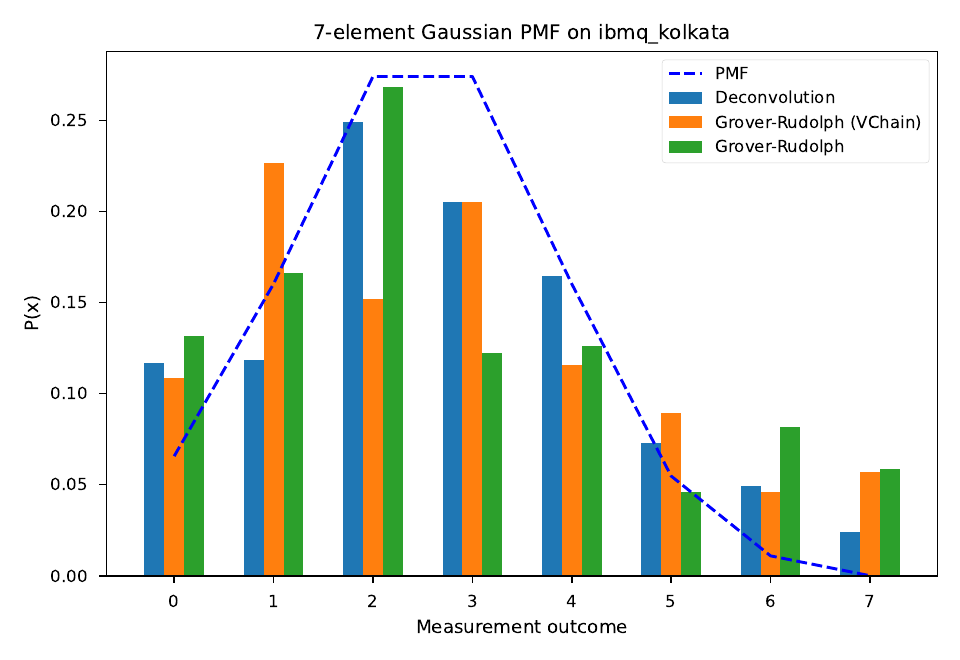} 
    \caption{7-element Gaussian PMF generated using IBMQ Kolkata (Table \ref{tab:7-gaussian-kolkata}).} 
    \label{fig:7-gaussian-kolkata} 
  \end{subfigure}
  \hfill
  \begin{subfigure}[b]{0.475\linewidth}
    \centering
    \includegraphics[width=\linewidth]{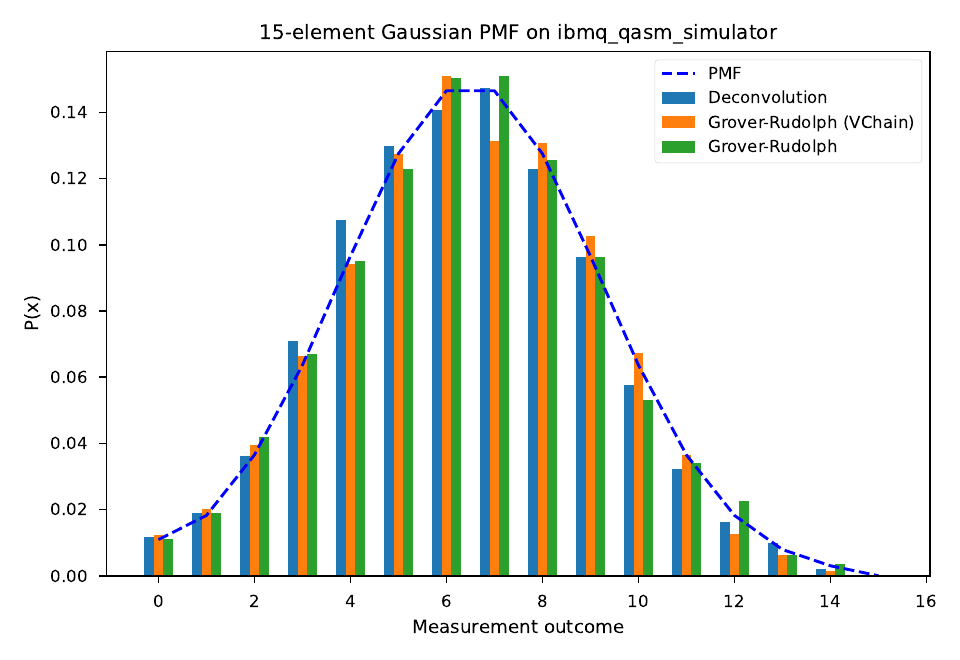} 
    \caption{15-element Gaussian PMF generated using IBMQ QASM simulator (Table \ref{tab:15-gaussian-sim}).} 
    \label{fig:15-gaussian-sim} 
  \end{subfigure}
  \hfill
  \begin{subfigure}[b]{0.475\linewidth}
    \centering
    \includegraphics[width=\linewidth]{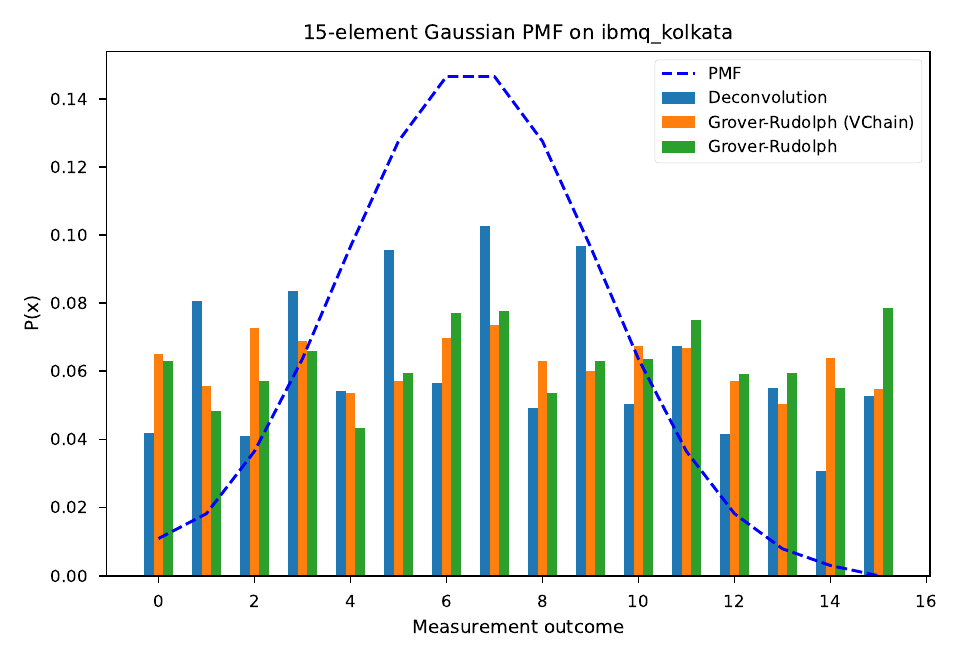} 
    \caption{15-element Gaussian PMF generated using IBMQ Kolkata (Table \ref{tab:15-gaussian-kolkata}).} 
    \label{fig:15-gaussian-kolkata} 
  \end{subfigure} 
  \caption{Discretized Gaussian PMF on a quantum computer: Fig.\ref{fig:7-gaussian-sim} and Fig. \ref{fig:15-gaussian-sim} show results on a noiseless simulator. The PMF is very well approximated by all methods, as seen in Table \ref{tab: gaussian}. 
  The novel deconvolution method produces the best results (lowest distance) with the fewest gates. 
  The same holds for the results on IBMQ Kolkata (Figures \ref{fig:7-gaussian-kolkata} and \ref{fig:15-gaussian-kolkata}), but it should be noted that noise limits the performance of the 15-element PMF.}
  \label{fig: gaussian} 
\end{figure}

\begin{figure}
  \centering
  \begin{subfigure}[b]{0.475\linewidth}
    \centering
    \includegraphics[width=\linewidth]{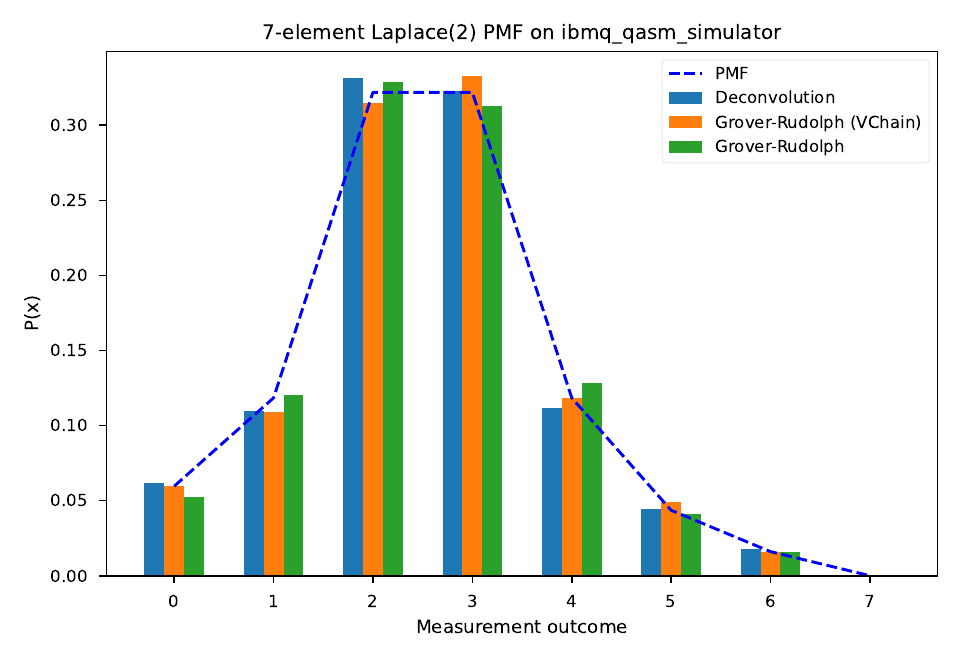} 
    \caption{7-element Laplace PMF generated using IBMQ QASM simulator (Table \ref{tab:7-laplace-sim}).} 
    \label{fig:7-laplace-sim} 
  \end{subfigure}
  \hfill
  \begin{subfigure}[b]{0.475\linewidth}
    \centering
    \includegraphics[width=\linewidth]{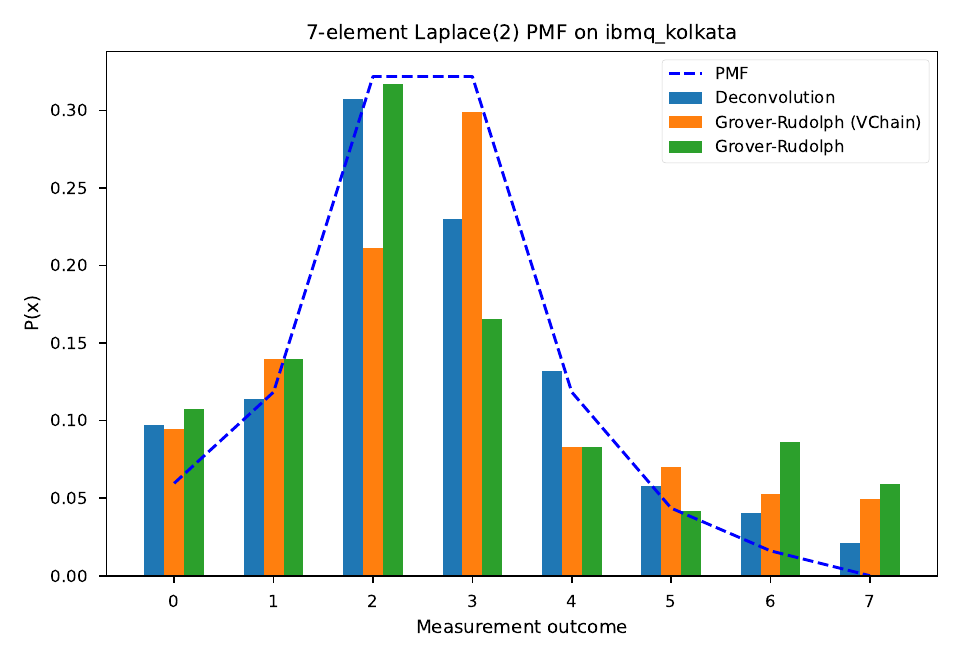} 
    \caption{7-element Laplace PMF generated using IBMQ Kolkata (Table \ref{tab:7-laplace-kolkata}).} 
    \label{fig:7-laplace-kolkata} 
  \end{subfigure}
  \hfill
  \begin{subfigure}[b]{0.475\linewidth}
    \centering
    \includegraphics[width=\linewidth]{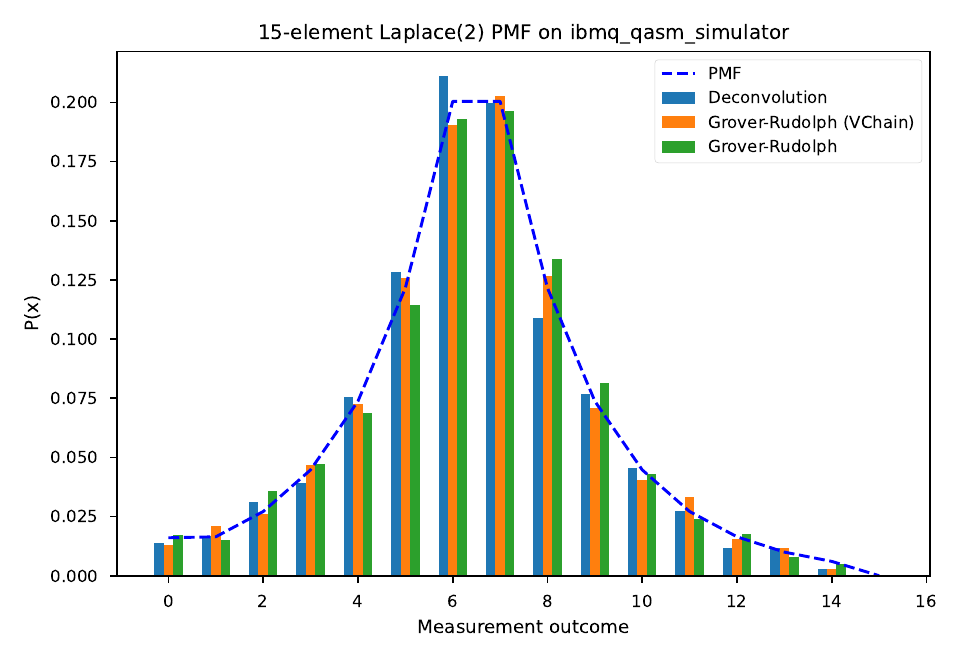} 
    \caption{15-element Laplace PMF generated using IBMQ QASM simulator (Table \ref{tab:15-laplace-sim}).} 
    \label{fig:15-laplace-sim} 
  \end{subfigure}
  \hfill
  \begin{subfigure}[b]{0.475\linewidth}
    \centering
    \includegraphics[width=\linewidth]{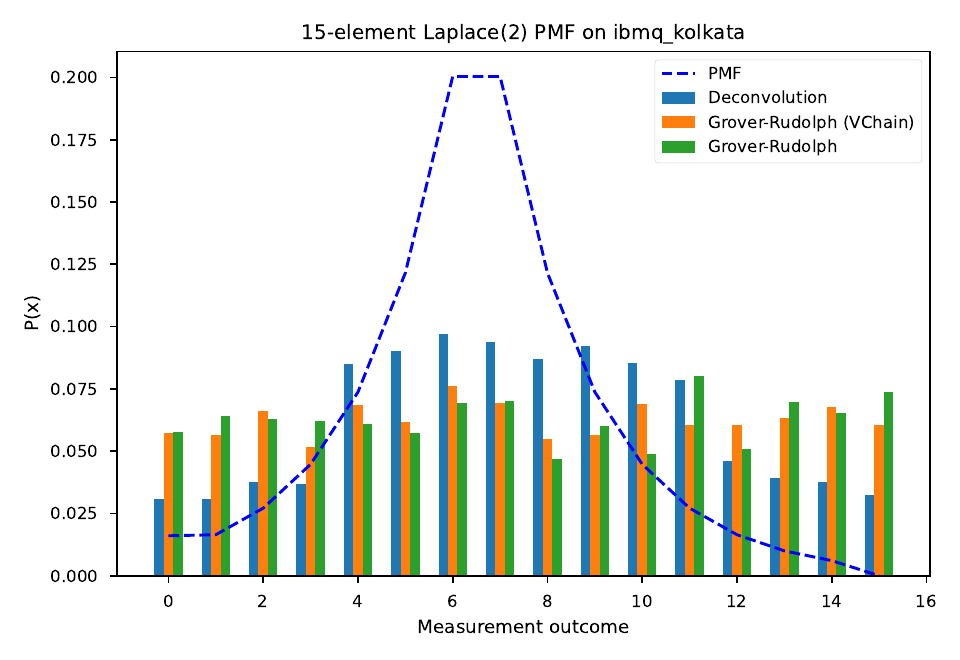} 
    \caption{15-element Laplace PMF generated using IBMQ Kolkata (Table \ref{tab:15-laplace-kolkata}).} 
    \label{fig:15-laplace-kolkata} 
  \end{subfigure} 
  \caption{Preparing a discretized Laplace $(\theta=2)$ PMF on a quantum computer (analogous to Fig. \ref{fig: gaussian}, but with other PMF). Figures \ref{fig:7-laplace-sim} and \ref{fig:15-laplace-sim} are the results on IBM's QASM simulator (noiseless). All methods very well approximate the PMFs.
  When looking at Table \ref{tab:laplace}, it is clear that the proposed deconvolution method results in the lowest distance with the lowest number of gates. When running on IBMQ Kolkata (figures \ref{fig:7-laplace-kolkata} and \ref{fig:15-laplace-kolkata}), the results are similar.}
  \label{fig: Laplacian} 
\end{figure}

Using the Qiskit primitive sampler, we can optimise the circuit and make it more resilient to noise \cite{runtime_sampler}.
We used optimisation level 3, which means the circuit is transpiled with 1Q gate optimisation, dynamical decoupling, commutative cancellation, and 2 qubit KAK optimisation.
We used resilience level 1, which means that errors associated with readout errors are mitigated with matrix-free measurement mitigation (M3). 
For more information on these techniques, we refer to qiskit runtime \cite{qiskit_runtime}.
The results for a Gaussian PMF are depicted in Fig. \ref{fig: gaussian} and Table \ref{tab: gaussian}, and the results for Laplacian PMF are shown in Fig. \ref{fig: Laplacian} and Table \ref{tab:laplace}.
It can be seen that the deconvolution method is able to approximate the PMF on the simulator well. 
Hence, compared to other methods, the deconvolution method works best on the quantum computer in terms of the QCV metric and circuit depth. 
Due to noise and circuit depth ($\mathcal{O}(10^2)$ for the 15-element PMF), the JS distance is of the order of $10^{-1}$ for all methods, but lowest for the deconvolution method. 
From Table \ref{tab: gaussian} and Fig. \ref{fig: gaussian}, it is clear that the deconvolution method has a shallower circuit and the best outcome compared to the GR state preparation method.
\begin{table}
  \begin{subtable}[t]{\linewidth}%
    \centering%
    \begin{tabular}{l||l|l|l|l}
        State preparation method & JS-distance                  & Circuit Depth & Active qubits & QCV     \\ \hline\hline      
        Deconvolution            & $\mathbf{8.84\cdot 10^{-4}}$ & \textbf{10}   & 6             & \textbf{60}        \\ \hline 
        Grover Rudolph (VChain)  & $1.93\cdot 10^{-3}$          & 30            & 4             & 120        \\ \hline         
        Grover Rudolph           & $4.99\cdot 10^{-4}$          & 33            & \textbf{3}    & 99                           
    \end{tabular}
    \caption{7-element Gaussian PMF on IBMQ QASM simulator (Fig. \ref{fig:7-gaussian-sim}).}
    \label{tab:7-gaussian-sim}
  \end{subtable}
  \vspace{0.5cm}
  \begin{subtable}[t]{\linewidth}
    \centering
    \begin{tabular}{l||l|l|l|l}
        State preparation method & JS-distance                  & Circuit Depth & Active qubits & QCV     \\ \hline\hline      
        Deconvolution            & $\mathbf{6.80\cdot 10^{-2}}$ & \textbf{54}   & 6             & \textbf{324}       \\ \hline 
        Grover Rudolph (VChain)  & $1.34\cdot 10^{-1}$          & 153           & \textbf{4}    & 612       \\ \hline          
        Grover Rudolph           & $1.66\cdot 10^{-1}$          & 134           & \textbf{4}    & 536                          
    \end{tabular}
    \caption{7-element Gaussian PMF on IBMQ Kolkata (Fig. \ref{fig:7-gaussian-kolkata}).}
    \label{tab:7-gaussian-kolkata}
  \end{subtable}
  \vspace{0.5cm}
  \begin{subtable}[t]{\linewidth}%
    \centering%
    \begin{tabular}{l||l|l|l|l}
        State preparation method & JS-distance                  & Circuit Depth & Active qubits & QCV\\ \hline\hline                   
        Deconvolution            & $\mathbf{1.69\cdot 10^{-3}}$ & \textbf{42}   & 11            & \textbf{462}               \\ \hline 
        Grover Rudolph (VChain)  & $2.37\cdot 10^{-3}$          & 104           & 6             & 624 \\ \hline                        
        Grover Rudolph           & $1.70\cdot 10^{-3}$          & 413           & \textbf{4}    & 1652                                 
    \end{tabular}
    \caption{15-element Gaussian PMF on IBMQ QASM simulator (Fig. \ref{fig:15-gaussian-sim}).}
    \label{tab:15-gaussian-sim}
  \end{subtable}
  \vspace{0.5cm}
  \begin{subtable}[t]{\linewidth}
    \centering
    \begin{tabular}{l||l|l|l|l}
        State preparation method & JS-distance                  & Circuit Depth & Active qubits & QCV\\ \hline\hline                    
        Deconvolved              & $\mathbf{2.35\cdot 10^{-1}}$ & \textbf{298}  & 11            & \textbf{3278}               \\ \hline 
        Grover Rudolph (VChain)  & $2.94\cdot 10^{-1}$          & 823           & \textbf{6}    & 4938     \\ \hline                    
        Grover Rudolph           & $3.13\cdot 10^{-1}$          & 796           & \textbf{6}    & 4776                                  
    \end{tabular}
    \caption{15-element Gaussian PMF on IBMQ Kolkata (Fig. \ref{fig:15-gaussian-kolkata}).}
    \label{tab:15-gaussian-kolkata}
  \end{subtable}
  \caption{Results of preparing Gaussian PMFs on the quantum simulators and IBMQ Kolkata. The JS distance is measured between the original discretized PMF and the measurement results. The deconvolution method has the lowest JS distance and circuit depth for all experiments.}
  \label{tab: gaussian}
\end{table}

To further ascertain the advantage of the deconvolution layer, we also present the results of loading Laplacian distribution using the deconvolution approach to show that the method is not only confined to the standard normal distribution but as described in Section \ref{sec: Motivation}, it is applicable for all bell-shaped distribution. 
The Laplacian distribution also comes under the log-concave probability distribution.
The probability density function corresponding to a random variable $\mathcal{X}$ that follows the Laplace distribution is \cite{kotz2001laplace}:
\begin{align} \label{eq: laplaceeq}
    f(x| \mu,\theta) = \frac{1}{2 \theta} e^{-\frac{|x - \mu|}{\theta}},
\end{align}
where the parameter $\mu$ is the mean value of the random variable $\mathcal{X}$, $\theta$ is called the scale parameter. 
$\mu \in \mathbb{R}$ is also called the location parameter,  whereas the scale parameter $\theta$ can only take positive real values i.e., $\theta > 0$.
We choose the parameters $\theta = 2$ and $\mu =0$ to validate our findings.

\begin{table}
  \begin{subtable}[t]{\linewidth}%
    \centering%
    \begin{tabular}{l||l|l|l|l}
        State preparation method & JS-distance                  & Circuit Depth & Active qubits & QCV \\ \hline\hline                 
        Deconvolution            & $\mathbf{5.88\cdot 10^{-4}}$ & \textbf{10}   & 6             & \textbf{60}               \\ \hline 
        Grover Rudolph (VChain)  & $7.25\cdot 10^{-4}$          & 30            & 4             & 120      \\ \hline                  
        Grover Rudolph           & $8.10\cdot 10^{-4}$          & 33            & \textbf{3}    & 99                                  
    \end{tabular}
    \caption{7-element Laplace PMF on IBMQ QASM simulator (Fig. \ref{fig:7-laplace-sim}).}
    \label{tab:7-laplace-sim}
  \end{subtable}
  \vspace{0.5cm}
  \begin{subtable}[t]{\linewidth}
    \centering
    \begin{tabular}{l||l|l|l|l}
        State preparation method & JS-distance                  & Circuit Depth & Active qubits & QCV \\ \hline\hline                   
        Deconvolution            & $\mathbf{4.88\cdot 10^{-2}}$ & \textbf{54}   & 6             & \textbf{324}                \\ \hline 
        Grover Rudolph (VChain)  & $9.76\cdot 10^{-2}$          & 159           & \textbf{4}    & 636       \\ \hline                   
        Grover Rudolph           & $1.50\cdot 10^{-1}$          & 133           & \textbf{4}    & 532                                   
    \end{tabular}
    \caption{7-element Laplace PMF on IBMQ Kolkata (Fig. \ref{fig:7-laplace-kolkata}).}
    \label{tab:7-laplace-kolkata}
  \end{subtable}
  \vspace{0.5cm}
  \begin{subtable}[t]{\linewidth}%
    \centering%
    \begin{tabular}{l||l|l|l|l}
        State preparation method & JS-distance                  & Circuit Depth & Active qubits & QCV \\ \hline\hline                  
        Deconvolution            & $\mathbf{3.00\cdot 10^{-3}}$ & \textbf{42}   & 11            & \textbf{462}               \\ \hline 
        Grover Rudolph (VChain)  & $2.53\cdot 10^{-3}$          & 104           & 6             & 624    \\ \hline                     
        Grover-Rudolph           & $2.56\cdot 10^{-3}$          & 413           & \textbf{4}    & 1652                                 
    \end{tabular}
    \caption{15-element Laplace PMF on IBMQ QASM simulator (Fig. \ref{fig:15-laplace-sim}).}
    \label{tab:15-laplace-sim}
  \end{subtable}
  \vspace{0.5cm}
  \begin{subtable}[t]{\linewidth}
    \centering
    \begin{tabular}{l||l|l|l|l}
        State preparation method & JS-distance                  & Circuit Depth & Active qubits & QCV \\ \hline\hline                  
        Deconvolution             & $\mathbf{1.75\cdot 10^{-1}}$ & \textbf{279}  & 11            & \textbf{3069}              \\ \hline 
        Grover-Rudolph (VChain)  & $3.40\cdot 10^{-1}$          & 815           & \textbf{6}    & 4890     \\ \hline                   
        Grover-Rudolph           & $3.76\cdot 10^{-1}$          & 798           & \textbf{6}    & 4788                                 
    \end{tabular}
    \caption{15-element Laplace PMF on IBMQ Kolkata (Fig. \ref{fig:15-laplace-kolkata}).}
    \label{tab:15-laplace-kolkata}
  \end{subtable}
  \caption{Results of preparing Laplace PMFs on the quantum simulator and IBMQ Kolkata. The JS distance is measured between the original discretized PMF and the measurement results. The deconvolution method has the lowest JS distance and lowest circuit depth for all experiments.}
  \label{tab:laplace}
\end{table}

\section{Conclusion and future work}
\label{sec: conclude}

In this article, we discussed a hybrid classical-quantum algorithm in which the classical step is based on the principle of deconvolution from signal processing. 
This classical preprocessing step helps in reducing the circuit depth required to load a given bell-shaped probability distribution. 
The deconvolution step breaks down the target PMF into smaller PMFs, which can be loaded in parallel into different quantum registers. 
The quantum part consists of (controlled) rotation gates followed by a quantum adder circuit, leading to an overall reduction in circuit depth. 
The algorithm reduces the dependency of the state preparation algorithms on big multi-controlled single-target gates used in traditional approaches for loading bell-shaped probability distributions.
We propose the deconvolution step discussed in this article as an additional classical step that can be merged with any existing state preparation algorithm for loading bell-shaped probability distributions, which will help in reducing the circuit depth further. 
Deconvolution is the inverse of convolution and is a kind of inverse problem which can be translated into a constrained optimization problem. 
We defined the JS function as the cost function for the optimization approach, and then we used the trust region method to find the optimum value of $\boldsymbol{q_1}$ and $\boldsymbol{q_2}$ that minimizes the cost function.
This deconvolution technique based on an optimization algorithm is developed as part of the proof of concept for our approach, and in the following section, we have discussed a more efficient deconvolution algorithm based on polynomial factorization. 
To verify state preparation using the deconvolution approach, we loaded 7 and 15-element PMF constructed by discretizing the two different probability distributions: (i) Standard Normal Distribution and (ii) Laplace distribution. 
The algorithm's results are positive on the QASM simulator and show a reduction in the circuit depth. 
Hence, the outcomes of the deconvolution method agree well with the expected outcome.
However, on the real hardware, more circuit depth reduction is still required since the noise is taking over.

We believe this work is the first where a theoretical concept from the signal processing field is used to solve the state preparation problem in quantum computation. 
Due to a shared foundation in the mathematical framework of probability and statistics. 
Signal processing and quantum computation have many overlapping and complementary topics. 
In future, we would like to dig further into different concepts of signal processing that can be used in quantum computation to design classical techniques that can efficiently load more complex problems into quantum processing units (QPUs) and provide us with more accurate results. 
We want to extend this method to solve the more complex problem of stochastic volatility modelling in finance, where the probability distribution is more complex. 
Calculations of some quantities of interest have no closed-form solutions, so they must be modelled using Monte Carlo simulation.
Once we can load the required probability distribution into the QPU with minimum circuit depth, we can use the remaining available time to calculate complex financial quantities. 
Furthermore, loading required probability distribution with minimum circuit depth into the current NISQ QPUs finds application in other fields. 
As mentioned earlier, an efficient solution to the loading problem increases the capability of quantum algorithms like QAE, HHL and quantum machine learning algorithms to solve complex problems. 

\section*{Reference}
\bibliographystyle{iopart-num}
\bibliography{ref}

\clearpage

\appendix
\section{Quantum Amplitude Estimation Algorithm}\label{sec: QAE_Algorithm}
The Quantum Amplitude Estimation (QAE) algorithm helps in estimating the value of `$a$' in the state $G_a\ket{0}_{{\mathfrak{n}}+1} = \sqrt{1-a} \ket{\psi_0}_{\mathfrak{n}}\ket{0} + \sqrt{a}\ket{\psi_1}_{\mathfrak{n}}\ket{1}$, where $G_a$ is an operator acting on $\ket{0}_{{\mathfrak{n}}+1}$ state. One major application of QAE is calculating the expectation value of functions with random variables as their input.
The calculation has three major steps, briefly outlined in the calculation VaR in Fig. \ref{fig:flow_chart}. But to re-emphasize and explain each step in detail, we outline them here more concretely.
\begin{itemize}
\step{Step 1} Load the required Probability distribution that the random variable follows into an $\mathfrak{n}$-qubit quantum register. 
For this, using an affine mapping, we can map the random variable $\mathcal{X}$ to the interval $\{0,1,2, \hdots, N-1\}$, where $N = 2^{\mathfrak{n}}$.  
Then, using any state preparation method discussed in this article, we can prepare a quantum operation that performs the following transformation 
\begin{align}
    R \ket{0}_{\mathfrak{n}} = \sum_{i=0}^{N-1}\sqrt{p_i}\ket{i}_{\mathfrak{n}} = \ket{\psi}_{\mathfrak{n}},
    \label{eq: QAE_Eq_1}
\end{align}
where $\ket{i}_n$ belongs to the computational basis set. 
In Equation \ref{eq: QAE_Eq_1}, $p_i$ represents the probability of superposition state $\ket{\psi}$ being in the basis state $\ket{i}_{\mathfrak{n}}$.
\step{Step 2} Load the function whose expectation is to be calculated using an extra qubit. 
We map the function to be calculated on the random variable to an operator $F$ that performs the following transformation
    \begin{align}
        F\ket{\psi}_{\mathfrak{n}}\ket{0} = \sum_{i=0}^{N-1}\sqrt{1 - f(i)}\sqrt{p_i}\ket{i}_{\mathfrak{n}}\ket{0} + \sum_{i=0}^{N-1}\sqrt{f(i)}\sqrt{p_i}\ket{i}_{\mathfrak{n}}\ket{1},
    \end{align}
    where $f(i)$ represents the function whose expectation we aim to calculate. 
    
    \step{Step 3} Use the QAE algorithm to estimate the expectation value of the function, which is nothing but the probability of measuring the last qubit in state $\ket{0}$. 
    For this estimation, the QAE algorithm requires only $O(M^{-1})$ samples, unlike $O(M^{-1/2})$ that other classical algorithms like the Monte Carlo require. 
\end{itemize}
Hence, the QAE algorithm gives a quadratic speed-up compared to the classical Monte Carlo algorithm. 
This advantage makes the QAE algorithm a very important algorithm for finance, particularly in the field of risk management.
\section{Grover Rudolph State Preparation Method} \label{app: GR_State_Preparation_Method}
In this section, we discuss the most commonly used state preparation method for loading log-concave probability distribution. 
Log-concave probability distributions, as explained earlier, are a family of probability distributions whose probability density functions are easily integrable using existing classical techniques \cite{grover2002creating}.
Let us assume that we want to prepare a given quantum register in a certain superposition state following some specific log-concave probability distribution $p(x)$.
In other words, we want to transform a quantum register that is initially in the state $\ket{0}_n$ to a superposition state $\ket{\psi}_n$ defined as 
\begin{align}
    \ket{\psi\left(\{p_i\}\right)} = \sum_{i}\sqrt{p_i}\ket{i}_{\mathfrak{n}},
    \label{eq: qubit_prob_state_1}
\end{align}
where $\ket{i}_n$ are computational basis states.
In the quantum framework, $\ket{i}$ is used to represent the values of a given random variable $\mathcal{X} \in \{0, 1, 2, \hdots, i, \hdots, 2^n-1\}$.
The probability of random variable $\mathcal{X}$ taking a value equal to $i$ is represented by $p_i$, which is equivalent to the probability of the quantum superposition state $\ket{\psi}$ collapsing into a state $\ket{i}$ when measured in computational basis states. 
Hence, we can create the superposition state as shown in Equation \ref{eq: qubit_prob_state_1} for a discretized version $\{p_i\}$ of $p(x)$.

\begin{figure}[h]
    \centering
    \includegraphics[scale = 0.25]{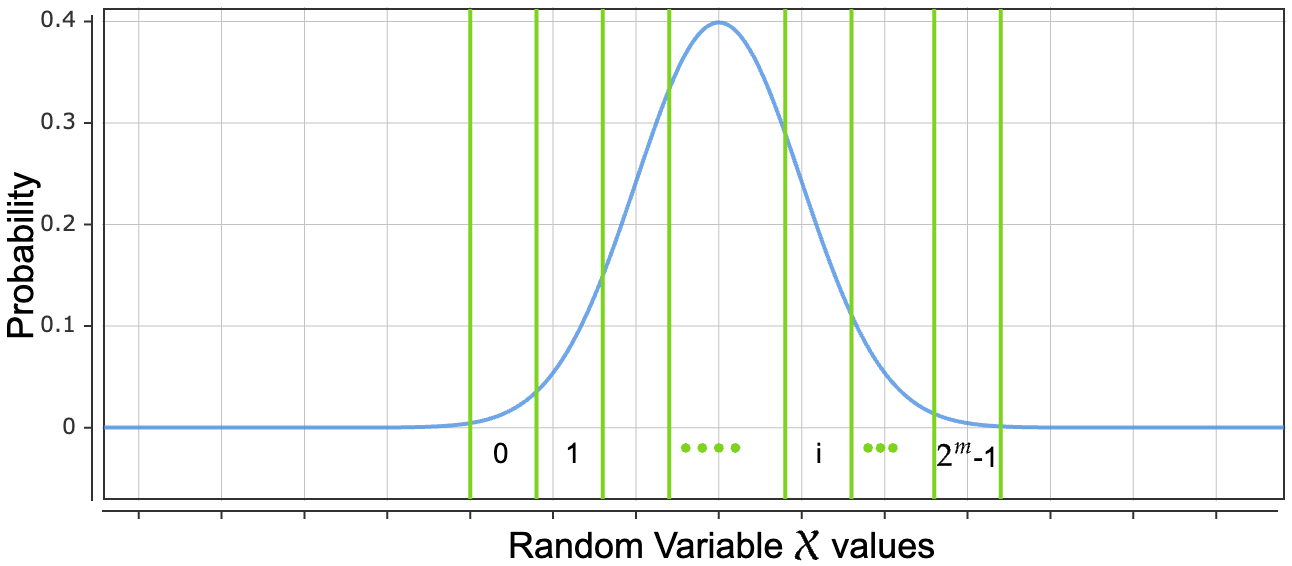}
    \caption{Normal distribution divided into $2^{\mathfrak{m}}$ regions. The $x$-axis of the plot represents the continuous values a random variable can take, and the $y$-axis represents the probability density of the random variable $\mathcal{X}$ taking each value. $p(x)$ of the normal distribution is given by $\frac{1}{\sigma \sqrt{2 \pi}}e^{-\frac{1}{2}(\frac{x-\mu}{\sigma})^2}$. In the context of plotting, we set the standard deviation value ($\sigma$) equal to $1$. However, it is important to note that we do not fix the value of the parameter $\mu$, which results in the absence of marking along the X-axis. Based on the value of $\mu$, the whole plot shifts right or left along the $x$-axis. }
    \label{fig: probability_distribution_division}
\end{figure}

Moving forward, we need to discretize the probability distribution corresponding to the random variable $\mathcal{X}$ over $N$ points, then the number of qubits needed to load this distribution is equal to $\mathfrak{n} =\left\lceil\log_2 N\right\rceil $.
This means, as shown in Fig. \ref{fig: probability_distribution_division}, if we have $\mathfrak{m}$ qubits, then we can divide the probability distribution into $2^\mathfrak{m}$ regions. 
The quantum superposition state of this $\mathfrak{m}$ qubits can be written as 
\begin{align}
    \ket{\psi}_\mathfrak{m} = \sum_{i = 0 }^{2^\mathfrak{m} - 1}\sqrt{p_i^{(\mathfrak{m})}} \ket{i},
\end{align}
where $p_i^{(\mathfrak{m})}$ represents the probability of the random variable $\mathcal{X}$ to lie in the $i^{th}$ region \cite{grover2002creating}. 
If we add one more qubit, we can subdivide this region further and have a total of $2^{\mathfrak{m}+1}$ regions. 
That is, we can further split $2^\mathfrak{m}$ regions into two halves, the left and right. 
Let the probability of random variable $\mathcal{X}$ to lie in the left and right half of the $i^{\mathrm{th}}$ region be $e_i$ and $f_i$, respectively. 
Then, using the $R_y$ rotation operator on the extra qubit controlled on the previous $\mathfrak{m}$ qubits, we can perform the following transformation 
\begin{align}
    \sum_{i=0}^{2^\mathfrak{m}-1} \sqrt{p_i^{(\mathfrak{m})}}\ket{i}\ket{0} = \sum_{i=0}^{2^\mathfrak{m}-2} \sqrt{p_i^{(\mathfrak{m})}}\ket{i}\ket{0} + \sqrt{p_{2^\mathfrak{m} - 1}^{(\mathfrak{m})}}\ket{2^\mathfrak{m}-1}\left(\sqrt{e_i}\ket{0} + \sqrt{f_i}\ket{1}\right).
\end{align}
We can do this for other regions and transform the state to a desired quantum state using a combination of $X$ and multi-controlled $R_y$ gates. 
Next, we repeat this step, increasing the qubit by one in each step.
We stop once the number of qubits reaches the desired value $\mathfrak{n}$. 
At the end of this iterative step, we would have constructed a circuit ($A$) consisting of $X$ and multi-controlled $R_y$ gates that would perform the following transformation
\begin{align}
    A\ket{0}_\mathfrak{n} = \sum_{i=0}^{2^\mathfrak{n}-1} \sqrt{p_i^{(\mathfrak{n})}}\ket{i}_\mathfrak{n}. 
\end{align}
If the start and end point of the $(2^\mathfrak{m} - 1)^{\mathrm{th}}$ region is marked as $x_{\mathrm{low}}$ and $x_{\mathrm{upp}}$ then we can calculate the value of $e_{2^\mathfrak{m}-1}$ and $f_{2^\mathfrak{m}-1}$ using the formula 
\begin{align}
    e_{2^\mathfrak{m}-1} &= \frac{\int_{x_{\mathrm{low}}}^{\frac{x_{\mathrm{upp}}-x_{\mathrm{low}}}{2}}p(x)}{\int_{x_\mathrm{low}}^{x_{\mathrm{upp}}}p(x)},\label{eq: calculation_cond_probability_1}\\ 
    f_{2^\mathfrak{m}-1} &= \frac{\int^{x_{\mathrm{upp}}}_{\frac{x_{\mathrm{upp}}-x_{\mathrm{low}}}{2}}p(x)}{\int_{x_\mathrm{low}}^{x_{\mathrm{upp}}}p(x)}.\label{eq: calculation_cond_probability_2}
\end{align}
The angle $\theta_i$ by which the last qubit must be rotated changes for each computational basis state $\ket{i}$ in the superposition state of  $\mathfrak{m}$ qubits ($\ket{\psi}_\mathfrak{m}$) and can be calculated using a simple mathematical equation 
\begin{align}
    \theta_i =  \cos^{-1}(e_i) \hspace{2mm} \text{or} \hspace{2mm} \theta_i =  \sin^{-1}(f_i).
    \label{eq: theta_value}
\end{align}
To illustrate this algorithm further, we explain it using an example in the following section. 
\subsection{Example for GR state preparation method}\label{sec: appendix_1_1}

\begin{figure}[h]
    \centering
    \begin{subfigure}{0.32\textwidth}  
        \includegraphics[width=\linewidth]{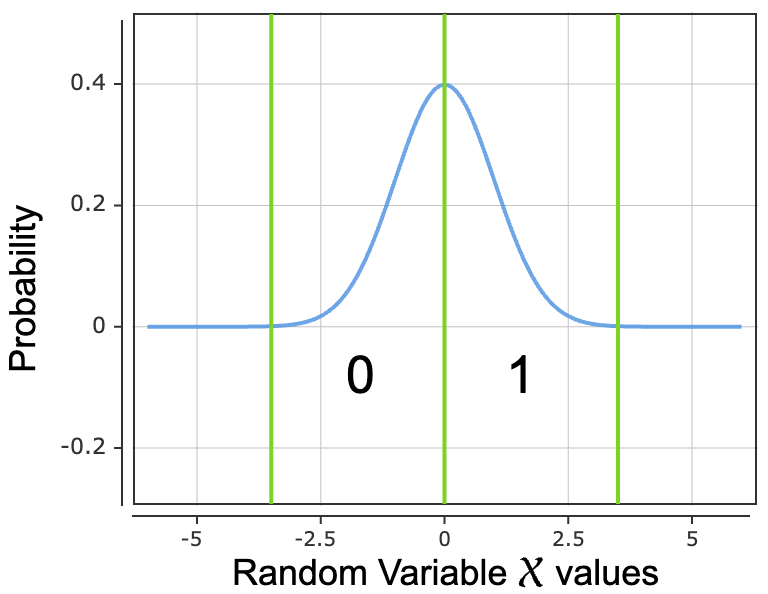}
        \caption{Standard Normal Distribution plot divided into 2 regions.}
        \label{fig: standard_normal_dsitribution_loading_1_a}
    \end{subfigure}
    \hfill
    \begin{subfigure}{0.32\textwidth}  
        \includegraphics[width=\linewidth]{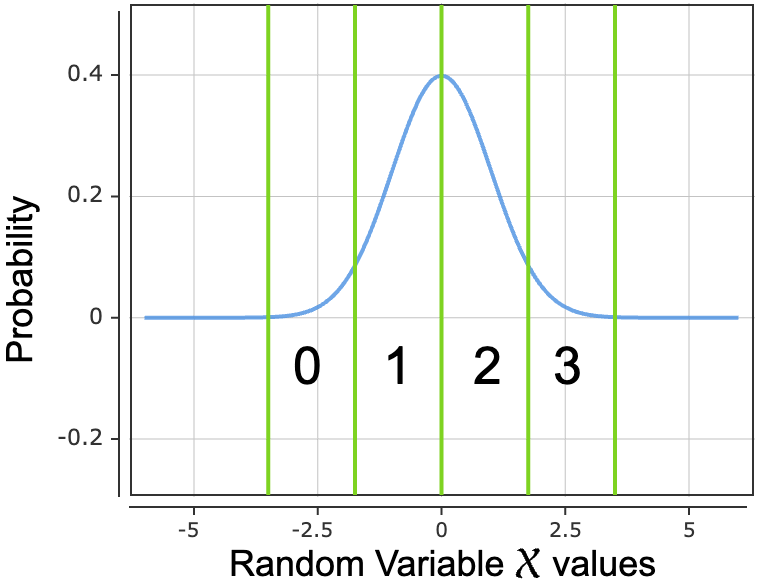}
       \caption{Standard Normal Distribution plot divided into 4 regions.}
        \label{fig: standard_normal_dsitribution_loading_1_b}
    \end{subfigure}
    \hfill
    \begin{subfigure}{0.32\textwidth}  
        \includegraphics[width=\linewidth]{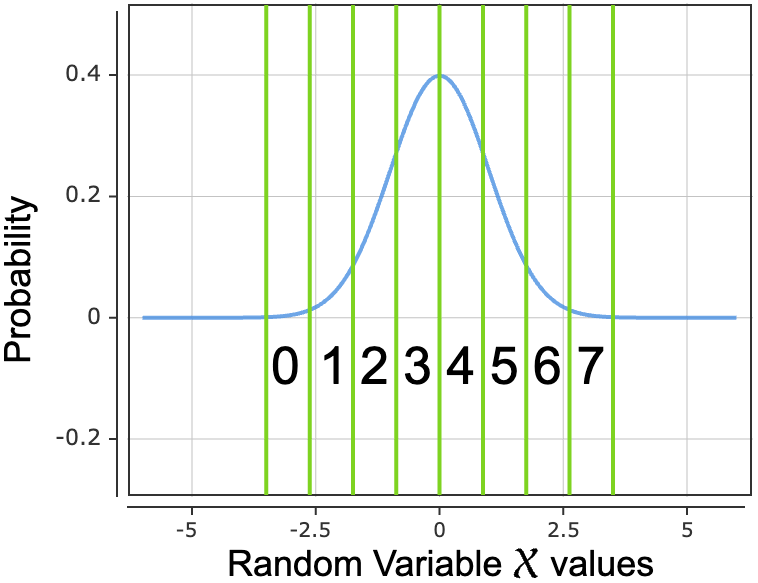}
        \caption{Standard Normal Distribution plot divided into 8 regions.}
        \label{fig:subfig_c}
    \end{subfigure}
    \caption{Standard Normal distribution plot subdivided into different regions as we recursively progress in the GR state preparation. }
    \label{fig: standard_normal_dsitribution_loading}
\end{figure}

Using the GR state preparation method, we demonstrate how to build a quantum circuit using multi-controlled $R_y$ and $X$ gates to load standard normal distribution into a quantum register consisting of $\mathfrak{n} = 3$ qubits.
The probability density function for standard normal distribution can be obtained by putting $\mu = 0$ and $\sigma = 1$ in the $p(x)$ formula corresponding to normal distribution.
Hence, $p(x)$ for normal distribution can be explicitly written as 
\begin{align}
    p(x) = \frac{1}{\sqrt{2\pi}}e^{\frac{-x^2}{2}},
\end{align}
where $x$ corresponds to different values random variable $\mathcal{X}$ can take along the $x$-axis. 
The plot for standard normal distribution is similar to the plot shown in Fig. \ref{fig: probability_distribution_division}, the only difference being the peak for the standard normal distribution will be located specifically at the origin since the value of the parameter $\mu$ equals to $0$ for standard normal distribution as shown in the Fig. \ref{fig: standard_normal_dsitribution_loading}.

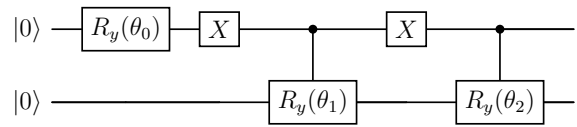
\begin{figure}[h]
    \centering
    \begin{tikzpicture}
    \node[scale = 0.8]{
    \begin{quantikz}
        \lstick{$\ket{0}$}  & \gate{R_y(\theta_0)} & \gate{X} & \ctrl{1}& \gate{X}&\ctrl{1}&\qw\\
        \lstick{$\ket{0}$}  & \qw & \qw  & \gate{R_y(\theta_1)}&\qw&\gate{R_y(\theta_2)}&\qw
    \end{quantikz}
    };
    \end{tikzpicture}
    \caption{Quantum circuit for loading a given probability distribution into a 2 qubit system using GR method.}
    \label{fig: Loading_2^2_state}
\end{figure}

We start by dividing the normal distribution into two regions, as shown in Fig. \ref{fig: standard_normal_dsitribution_loading_1_a} and the load the probability of being in the left half or right half as the probability of the first qubit being in the state $\ket{0}$ or $\ket{1}$ respectively. 
Next, we further subdivide this region into two halves, resulting in a total of four regions. The probability of the random variable being in the left half of the region marked zero in Fig. \ref{fig: standard_normal_dsitribution_loading_1_a}, which is the same as the region marked zero in Fig. \ref{fig: standard_normal_dsitribution_loading_1_b}, is equal to the probability of second qubit being in state $\ket{0}$ given that the first qubit is in state $\ket{0}$.
Similarly, we can define the probability of the first and second qubits being in the state $\ket{01}$,$\ket{10}$ and $\ket{11}$. 
The quantum circuit constructed following the GR state preparation method to prepare the first and second qubits in this state is shown in Fig. \ref{fig: Loading_2^2_state}.

The values of $\theta_0$, $\theta_1$ and $\theta_2$ can be calculated using the formal given in Equation \ref{eq: calculation_cond_probability_1}, \ref{eq: calculation_cond_probability_2} and \ref{eq: theta_value}. 
For $\mathfrak{n}=3$, we repeat this step just one more time and for any arbitrary value of $\mathfrak{n}$, we repeat the step until the number of regions is equal to $2^\mathfrak{n}$. 
Fig. \ref{fig: 3_qubit_state_preparation} shows the quantum circuit to load the discretized version of the normal distribution.  

\begin{figure}[h]
    \centering
    \begin{tikzpicture}
    \node[scale=0.73]{
    \begin{quantikz}[column sep=1.3mm]
        \lstick{$\ket{0}$}&\gate{R_y(\theta_0)} &\gate{X} &\ctrl{1} &\gate{X} &\ctrl{1} &\gate{X} &\ctrl{1} &\gate{X} &\gate{X}  &\ctrl{1}&\gate{X}&\ctrl{1}&\qw&\ctrl{1} &\qw\\
        \lstick{$\ket{0}$}&\qw &\qw &\gate{R_y(\theta_1)}&\qw &\gate{R_y(\theta_2)} &\gate{X} &\ctrl{1} &\gate{X} &\qw   &\ctrl{1}&\gate{X}&\ctrl{1}&\gate{X} &\ctrl{1} &\qw\\
        \lstick{$\ket{0}$} &\qw &\qw &\qw &\qw &\qw &\qw &\gate{R_y(\theta_3)} &\qw &\qw &\gate{R_y(\theta_4)} &\qw  &\gate{R_y(\theta_5)} 
        &\qw &\gate{R_y(\theta_6)} &\qw
    \end{quantikz}
    };
    \end{tikzpicture}
    \caption{Quantum circuit built using GR method for loading required probability distribution into a 3 qubit system. }
    \label{fig: 3_qubit_state_preparation}
\end{figure}
In Fig. \ref{fig: 3_qubit_state_preparation}, we can calculate the values of the $\theta$'s by following the Equations given in \ref{eq: calculation_cond_probability_1}, \ref{eq: calculation_cond_probability_2} and \ref{eq: theta_value}.
\section{VChain}
\label{sec: VChain}
In this section, we briefly discuss the VChain implementation of a multi-controlled single target gate. 
If the gate has $\mathfrak{n}$ control qubits, then we need $\mathfrak{n}-1$ ancilla qubits for the VChain implementation of the gate. 
As shown in Figure \ref{fig: vchain}, we use Toffoli gates to flip the ancilla qubits one by one based on the value of the control qubits and at last, we apply the controlled $R_y$ gate on the target qubit based on the value of the last ancilla qubit. 
All the ancilla qubits are reset to state $\ket{0}$ by reversing the pattern in which we applied the Toffoli gates.
This forms a V-shaped pattern of Toffoli gates, and hence the method is called VChain. 
\begin{figure}[h]
    \centering
    \footnotesize
    \begin{tikzpicture}
    \node[scale=0.85]{
    \begin{quantikz}[row sep={0.6cm,between origins}, align equals at=1.5, column sep=0.3cm]
         & \lstick{${q}_{0}$} & \ctrl{1} & \qw & & &
         \lstick{${q}_{0}$} & \ctrl{1} & \qw & \qw & \qw & \qw & \qw & \qw & \qw & \ctrl{1} & \qw & \qw\\
    	 & \lstick{${q}_{1}$} & \ctrl{1} & \qw & & &
         \lstick{${q}_{1}$} & \ctrl{5} & \qw & \qw & \qw & \qw & \qw & \qw & \qw & \ctrl{5} & \qw & \qw\\
	 	 & \lstick{${q}_{2}$} & \ctrl{1} & \qw & & &
         \lstick{${q}_{2}$} & \qw & \ctrl{4} & \qw & \qw & \qw & \qw & \qw & \ctrl{4} & \qw & \qw & \qw\\
	 	 & \lstick{${q}_{3}$} & \ctrl{1} & \qw & & &
         \lstick{${q}_{3}$} & \qw & \qw & \ctrl{4} & \qw & \qw & \qw & \ctrl{4} & \qw & \qw & \qw & \qw\\
	 	 & \lstick{${q}_{4}$} & \ctrl{1} & \qw & \hfill = \hfill & &
         \lstick{${q}_{4}$} & \qw & \qw & \qw & \ctrl{4} & \qw & \ctrl{4} & \qw & \qw & \qw & \qw & \qw\\
	 	 & \lstick{${q}_{5}$} & \gate{R_y\,(\mathrm{\theta})} & \qw & & &
         \lstick{${q}_{5}$} & \qw & \qw & \qw & \qw & \gate{R_y\,(\mathrm{\theta})} & \qw & \qw & \qw & \qw & \qw & \qw\\
	 	 & & & & & & \lstick{${a}_{0}$} & \targ{} & \ctrl{1} & \qw & \qw & \qw & \qw & \qw & \ctrl{1} & \targ{} & \qw & \qw\\
         & & & & & & \lstick{${a}_{1}$} & \qw & \targ{} & \ctrl{1} & \qw & \qw & \qw & \ctrl{1} & \targ{} & \qw & \qw & \qw\\
	 	 & & & & & & \lstick{${a}_{2}$} & \qw & \qw & \targ{} & \ctrl{1} & \qw & \ctrl{1} & \targ{} & \qw & \qw & \qw & \qw\\
	 	 & & & & & & \lstick{${a}_{3}$} & \qw & \qw & \qw & \targ{} & \ctrl{-4} & \targ{} & \qw & \qw & \qw & \qw & \qw\\
     \end{quantikz}
     };
     \end{tikzpicture}
 \caption{V-Chain structure to create a multi-controlled $R_y(\theta)$-gate. The multi-controlled $R_y(\theta)$-gate does not exist on current hardware and thus will need to be decomposed. The decomposition in basis gates without ancilla qubits will have more circuit depth compared to the decomposition in which we use ancilla qubits \cite{MCMT}. 
 Using ancilla qubits and Toffoli gates to create the $R_y(\theta)$-gate will result in a linear-depth decomposed circuit at the expense of adding qubits. One can visually verify the two circuits are equivalent by considering the state of $q_0, q_1, ..., q_4$.
 The $R_y(\theta)$-gate will only be applied with input $\ket{11111}$, as should be the case for a multi-controlled gate.}
 \label{fig: vchain}
 \end{figure}
 \section{Gradient and Hessian matrix calculation for deconvolution using Trust region method}
 \label{sec: Gradient_Hessian_fn_calculation}
 The gradient of the JS function can be calculated using the formula
\begin{align}
    \label{eq : gradientequation}
    \frac{\partial JS(\boldsymbol{P}||\boldsymbol{Q})}{\partial q_{1_{k}}} &= \sum_{i=0}^{L(\boldsymbol{q_1})+L(\boldsymbol{q_2})-1}  \frac{\partial Q_i}{\partial q_{1_{k}}}\log\left(\frac{Q_i}{R_i}\right),
\end{align}
where, 
\begin{align}
    Q_i = \sum_{u = \max\left(0,i-L(\boldsymbol{q_1})+1\right)}^{\min\left(i,L(\boldsymbol{q_2})-1\right)} q_{1_u} q_{2_{i-u}} &\implies
    \frac{\partial Q_i}{\partial q_{1_k}} = \sum_{u = \max\left(0,i-L(\boldsymbol{q_1})+1\right)}^{\min\left(i,L(\boldsymbol{q_2})-1\right)} q_{2_{i-u}} \frac{\partial q_{1_u}}{q_{1_k}}.
\end{align}
In equation \ref{eq : gradientequation} $R_i$ represents the $i^{\mathrm{th}}$ element of the PMF $\boldsymbol{R}$ defined using the equation \ref{eq: Rdef}.
In the equations \ref{eq: hes1}, \ref{eq : hes3}, \ref{eq: hes4} and \ref{eq:hes2}, the index $i$ runs from $i=0 $ to $i = L(\boldsymbol{q_1}) + L(\boldsymbol{q_2}) -1$ and the index $u$ runs from $ u = \max(0,i-L(\boldsymbol{q_1})+1) $ to $ \min(i,L(\boldsymbol{q_2})-1)$.
 The second-order derivative for the JS distance function is given by
\begin{align}
    A_{lk} = \frac{\partial JS(\boldsymbol{P}||\boldsymbol{Q})}{\partial q_{1_l}\partial q_{1_k}} &= \sum_i \left(\left(\frac{1}{Q_i} - \frac{1}{2 R_i}\right)\left(\sum_u q_{2_{i-u}}\frac{\partial q_{1_u}}{\partial q_{1_k}}\right)\left(\sum_u q_{2_{i-u}}\frac{\partial q_{1_u}}{\partial q_{1_l}}\right)\right), \label{eq: hes1}\\
    \begin{split}
       B_{lk} = \frac{\partial JS(\boldsymbol{P}||\boldsymbol{Q})}{\partial q_{2_l}\partial q_{1_k}} &= \sum_i \log \frac{Q_i}{R_i}\left(\sum_u \frac{\partial q_{2_{i-u}}}{\partial q_{2_l}}\frac{\partial q_{1_u}}{\partial q_{1_k}}\right) + \\
        & \hspace{5mm} \sum_i \left(\left(\frac{1}{Q_i} - \frac{1}{2 R_i}\right)\left(\sum_u q_{2_{i-u}}\frac{\partial q_{1_u}}{\partial q_{1_k}}\right)\left(\sum_u q_{1_u} \frac{\partial q_{2_{i-u}}}{\partial q_{2_l}}\right)\right),
     \end{split}\label{eq : hes3}\\
     C_{lk} = \frac{\partial JS(\boldsymbol{P}||\boldsymbol{Q})}{\partial q_{2_l}\partial q_{2_k}} &= \sum_i \left(\left(\frac{1}{Q_i} - \frac{1}{2 R_i}\right)\left(\sum_u q_{1_u}\frac{\partial q_{2_{i-u}}}{\partial q_{2_k}}\right)\left(\sum_u q_{1_u}\frac{\partial q_{2_{i-u}}}{\partial q_{2_l}}\right)\right),\label{eq: hes4}
\end{align}
\begin{align}
     \begin{split}
        D_{lk}= \frac{\partial JS(\boldsymbol{P}||\boldsymbol{Q})}{\partial q_{1_l}\partial q_{2_k}} &= \sum_i \log \frac{Q_i}{R_i}\left(\sum_u \frac{\partial q_{1_{u}}}{\partial q_{1_l}}\frac{\partial q_{2_{i-u}}}{\partial q_{2_k}}\right) + \\
        & \hspace{5mm} \sum_i \left(\left(\frac{1}{Q_i} - \frac{1}{2 R_i}\right)\left(\sum_u q_{2_{i-u}}\frac{\partial q_{1_u}}{\partial q_{1_l}}\right)\left(\sum_u q_{1_u} \frac{\partial q_{2_{i-u}}}{\partial q_{2_k}}\right)\right).
     \end{split} 
     \label{eq:hes2}
\end{align}
The Hessian matrix, can be constructed using the equations \ref{eq: hes1}, \ref{eq : hes3}, \ref{eq: hes4} and \ref{eq:hes2} as given below :
\begin{align}
    H_{(m+n) \times (m+n)}=\begin{pmatrix}
\begin{array}{c|c}
     A_{m\times m}&D_{m\times n} \\
     ---&--- \\
      B_{n \times m}&C_{n \times n} 
\end{array}\end{pmatrix}.
\label{eq: matrix_general_form }
\end{align}
\section{Example for Algorithm 3}
\label{sec: appendix_2}
In this Section, we take the example of a PMF $\boldsymbol{P}$ that we obtain by discretizing the normal distribution over $2^5$ points.
This implies the length (number of elements) in the PMF vector $\boldsymbol{P}$ is $32$ and the PGF $f(x)$ obtained using the Equation \ref{eq: pgf_calculation}  corresponds to a polynomial of degree $L(\boldsymbol{P}) - 1 = 32 - 1 = 31$,
\begin{align}
    f(x) = & 0.001111 + 0.00187962x + 0.00307045x^2 + 0.00484294x^3 +
       0.00737552x^4 + 0.01084556x^5 \nonumber \\ 
       & + 0.01539884x^6 + 0.02111057x^7 + 0.02794398x^8 + 0.0357152x^9 + 0.04407519x^{10} \nonumber\\ 
       & + 0.05251843x^{11} + 0.06042348x^{12} + 0.06712375x^{13}+ 0.07199846x^{14} + 0.074567x^{15}\nonumber \\ 
       & + 0.074567x^{16} + 0.07199846x^{17} + 0.06712375x^{18} + 0.06042348x^{19} + 0.05251843x^{20} \nonumber \\ 
       & + 0.04407519x^{21} + 0.0357152x^{22} + 0.02794398x^{23} + 0.02111057x^{24} + 0.01539884x^{25} \nonumber \\
       & + 0.01084556x^{26} + 0.00737552x^{27} + 0.00484294x^{28} + 0.00307045x^{29} + 0.00187962x^{30} \nonumber \\
       &+ 0.001111x^{31}.
\end{align}
The first step of Algorithm \ref{alg: Poly_deconv} is to calculate all the roots of the polynomial $f(x)$ and then segregate them into three categories (i)  Complex roots with non-negative real,(ii) Complex roots with a negative real part and (iii) real roots. 
Then, we arrange the category (i) roots in ascending order based on the real part and name it $basket_1$.
$basket_1$ is a list of 1D arrays, where the 1D arrays containing the conjugate pair and for this particular example 
\begin{align}
    basket_1 = [&[\num[round-mode=places, round-precision=5]{0.1750446257566166}+\num[round-mode=places, round-precision=5]{0.9845605004232698}j, \num[round-mode=places, round-precision=5]{0.17504462575661892}-\num[round-mode=places, round-precision=5]{0.9845605004232723}j],[\num[round-mode=places, round-precision=5]{0.3597590272791502}-\num[round-mode=places, round-precision=5]{0.9330452520061215}j, \num[round-mode=places, round-precision=5]{0.359759027279154}+\num[round-mode=places, round-precision=5]{0.9330452520061235}j], \nonumber \\
                &[\num[round-mode=places, round-precision=5]{0.5302711299415993}+\num[round-mode=places, round-precision=5]{0.8478281245337772}j, \num[round-mode=places, round-precision=5]{0.5302711299415998}-\num[round-mode=places, round-precision=5]{0.8478281245337765}j], [\num[round-mode=places, round-precision=5]{0.682088238167622}-\num[round-mode=places, round-precision=5]{0.731269878603911}j, \num[round-mode=places, round-precision=5]{0.6820882381676232}+\num[round-mode=places, round-precision=5]{0.7312698786039119}j], \nonumber \\
                &[\num[round-mode=places, round-precision=5]{0.7059889885210886}-\num[round-mode=places, round-precision=5]{0.41031265564727076}j, \num[round-mode=places, round-precision=5]{0.7059889885210903}+\num[round-mode=places, round-precision=5]{0.4103126556472708}j], [\num[round-mode=places, round-precision=5]{0.776903022928539}-\num[round-mode=places, round-precision=5]{0.6296202768053927}j, \num[round-mode=places, round-precision=5]{0.7769030229285399}+\num[round-mode=places, round-precision=5]{0.6296202768053933}j], \nonumber \\
                &[\num[round-mode=places, round-precision=5]{1.058808365464415}+\num[round-mode=places, round-precision=5]{0.6153672073063402}j, \num[round-mode=places, round-precision=5]{1.0588083654644178}-\num[round-mode=places, round-precision=5]{0.6153672073063433}j]].
\end{align}
Also, merge the categories (ii) and (iii) roots and store them in a list variable $basket_2$ and for this  particular example 
\begin{align}
    basket_2 = [&[\num[round-mode=places, round-precision=5]{-0.016895893550830186}+\num[round-mode=places, round-precision=5]{0.9998572542023804}j, \num[round-mode=places, round-precision=5]{-0.016895893550830644}-\num[round-mode=places, round-precision=5]{0.9998572542023763}j], \nonumber \\
                &[\num[round-mode=places, round-precision=5]{-0.20875549693350764}-\num[round-mode=places, round-precision=5]{0.9779678637358403}j, \num[round-mode=places, round-precision=5]{-0.20875549693350828}+\num[round-mode=places, round-precision=5]{0.9779678637358418}j], \nonumber \\
                &[-\num[round-mode=places, round-precision=5]{0.393198331956103}+\num[round-mode=places, round-precision=5]{0.9194536811318694}j, -\num[round-mode=places, round-precision=5]{0.3931983319561042}-\num[round-mode=places, round-precision=5]{0.9194536811318671}j], \nonumber \\
                &[\num[round-mode=places, round-precision=5]{-0.563157764980587}+\num[round-mode=places, round-precision=5]{0.8263494005213874}j, \num[round-mode=places, round-precision=5]{-0.5631577649805874}\num[round-mode=places, round-precision=5]{-0.8263494005213892}j], \nonumber \\
                &[\num[round-mode=places, round-precision=5]{-0.7121153156206508}+\num[round-mode=places, round-precision=5]{0.7020625166311788}j, \num[round-mode=places, round-precision=5]{-0.7121153156206516}\num[round-mode=places, round-precision=5]{-0.7020625166311799}j], \nonumber \\ 
                &[\num[round-mode=places, round-precision=5]{-0.8343547541714113}+\num[round-mode=places, round-precision=5]{0.551227851429491}j, \num[round-mode=places, round-precision=5]{-0.8343547541714125}-\num[round-mode=places, round-precision=5]{0.5512278514294874}j], \nonumber \\ 
                &[\num[round-mode=places, round-precision=5]{-0.9251836312976823}-\num[round-mode=places, round-precision=5]{0.379519760195477}j, \num[round-mode=places, round-precision=5]{-0.9251836312976831}+\num[round-mode=places, round-precision=5]{0.37951976019547573}j], \nonumber \\
                &[\num[round-mode=places, round-precision=5]{-0.9811146055103414}+\num[round-mode=places, round-precision=5]{0.1934273270618093}j), \num[round-mode=places, round-precision=5]{-0.9811146055103458}-\num[round-mode=places, round-precision=5]{0.19342732706180849}j], \nonumber \\
                &[\num[round-mode=places, round-precision=5]{-1.0000000000000024}+\num[round-mode=places, round-precision=5]{0}j]].
\end{align}
$basket_2$ list is randomly shuffled, and after this step, we start a while loop as shown in Fig. \ref{fig: Flowchart_Deconvl_Algorithm}.
In the first iteration, we take the first element from $basket_1$ which is a 1D array $[\num[round-mode=places, round-precision=5]{0.1750446257566166}+\num[round-mode=places, round-precision=5]{0.9845605004232698}j, \num[round-mode=places, round-precision=5]{0.17504462575661892}-\num[round-mode=places, round-precision=5]{0.9845605004232723}j]$.
Then we start another while, in which we select a random element from $basket_2$ lets say $[\num[round-mode=places, round-precision=5]{-0.563157764980587}+\num[round-mode=places, round-precision=5]{0.8263494005213874}j,$ $\num[round-mode=places, round-precision=5]{-0.5631577649805874}-\num[round-mode=places, round-precision=5]{0.8263494005213892}j]$ and both the element into one single 1D array, i.e.,
\begin{align}
    \text{temp\_array} = [&\num[round-mode=places, round-precision=5]{0.1750446257566166}+\num[round-mode=places, round-precision=5]{0.9845605004232698}j, \num[round-mode=places, round-precision=5]{0.17504462575661892}-\num[round-mode=places, round-precision=5]{0.9845605004232723}j, \num[round-mode=places, round-precision=5]{-0.563157764980587}+\num[round-mode=places, round-precision=5]{0.8263494005213874}j,\nonumber \\ &\num[round-mode=places, round-precision=5]{-0.5631577649805874}-\num[round-mode=places, round-precision=5]{0.8263494005213892}j].
\end{align}
Using the Python function numpy.polynomial.polyfromroots, we calculate a monic polynomial that has roots mentioned in temp\_array.
We delete the random element we choose from the $basket_2$ list and check if all the coefficients of the monic polynomial are greater than zero. 
If yes, we terminate the current while loop and append the temp\_array to the $basket_2$ list and also delete the element we choose from $basket_1$ from $baket_1$ itself.
Otherwise, we select another random element and repeat the whole procedure. 
The outer while loop ends when we have exhausted all the elements from $baslet_1$.
We run Algorithm \ref{alg: Poly_deconv} in parallel in many cores and select the result which would have factorised the target polynomial $f(x)$ into the most number of factors. 
Among this result, we select the one whose biggest factor has the least degree compared to other biggest factors belonging to other results.  
For our specific example using Algorithm \ref{alg: Poly_deconv}, we get the following result

\begin{align}
    f(x) = & \left(1.0 + \num[round-mode=places, round-precision=5]{0.77622628}x + \num[round-mode=places, round-precision=5]{1.60568904}x^2 + \num[round-mode=places, round-precision=5]{0.77622628}x^3 + 1.0x^4\right) \left(1.0 + \num[round-mode=places, round-precision=5]{0.06005415}x + \num[round-mode=places, round-precision=5]{0.05709808}x^2 \right.\nonumber \\
    &\left.+ \num[round-mode=places, round-precision=5]{0.06005415}x^3 + 1.0x^4\right) \left(1.0 + \num[round-mode=places, round-precision=5]{0.65352995}x + \num[round-mode=places, round-precision=5]{0.14493192}x^2 + \num[round-mode=places, round-precision=5]{1.16946487}x^3 + \num[round-mode=places, round-precision=5]{0.93393201}x^4 \right.\nonumber \\
    & \left. + \num[round-mode=places, round-precision=5]{0.93393201}x^5 + \num[round-mode=places, round-precision=5]{1.16946487}x^6 + \num[round-mode=places, round-precision=5]{0.14493192}x^7 + \num[round-mode=places, round-precision=5]{0.65352995}x^8 + 1.0x^9\right)\left(1.0 + \num[round-mode=places, round-precision=5]{0.20201441}x \right. \nonumber\\
    & \left. + \num[round-mode=places, round-precision=5]{0.06184121}x^2 + \num[round-mode=places, round-precision=5]{0.39708601}x^3 + \num[round-mode=places, round-precision=5]{0.23166121}x^4 + \num[round-mode=places, round-precision=5]{0.98690859}x^5 + \num[round-mode=places, round-precision=5]{1.56270069}x^6 + \num[round-mode=places, round-precision=5]{1.38605043}x^7\right. \nonumber\\
    & \left. + \num[round-mode=places, round-precision=5]{1.56270069}x^8 + \num[round-mode=places, round-precision=5]{0.98690859}x^9 + \num[round-mode=places, round-precision=5]{0.23166121}x^{10}+ \num[round-mode=places, round-precision=5]{0.39708601}x^{11}+ \num[round-mode=places, round-precision=5]{0.06184121}x^{12} + \num[round-mode=places, round-precision=5]{0.20201441}x^{13} \right. \nonumber \\
    & \left. + 1.0x^{14}\right).
\end{align}
For this example, Algorithm \ref{alg: Poly_deconv} ran over $1000$ runs, which were parallelized. 
The result was obtained in $50s$, which is much faster compared to the deconvolution algorithm developed using optimization techniques in Section \ref{sec: trust_region}.

\end{document}